\documentclass[showpacs,amsmath,amssymb,aps,pre,twocolumn,]{revtex4-1}
\usepackage{graphicx}
\usepackage{dcolumn}
\usepackage{bm}
\usepackage[breaklinks,colorlinks = true,linkcolor = red,urlcolor=blue,citecolor=red]{hyperref}
\usepackage{multirow}
\usepackage{array}
\usepackage{booktabs}
\usepackage{ctable}
\usepackage{ulem}
\usepackage{upgreek}
\usepackage{epsfig}
\usepackage{mathrsfs}
\usepackage{amssymb}
\usepackage{amsbsy}
\usepackage{color}
\usepackage{cancel}
\usepackage{pifont}
\usepackage{marginnote}
\usepackage{float}
\usepackage{verbatim}
\usepackage{subfigure}
\usepackage{bbm, dsfont}
\usepackage{lineno}
\usepackage{amsmath}
\definecolor{dgreen}{rgb}{0,0.7,0}

\def\greenw#1{{\color{black} #1}}
\def\bluew#1{{\color{black} #1}}

\usepackage{xr}
\makeatletter

\newcommand{\Rmnum}[1]{\expandafter\@slowromancap\romannumeral #1@}
\makeatother

\usepackage[utf8]{inputenc}

\begin{abstract}
Heterogeneous diffusion with spatially changing diffusion coefficient arises in many experimental systems like protein dynamics in the cell cytoplasm, mobility of cajal bodies and confined hard-sphere fluids. Here, we showcase a simple model of heterogeneous diffusion where the diffusion coefficient $D(x)$ varies in power-law way, i.e. $D(x) \sim |x|^{-\alpha}$ with the exponent $\alpha >-1$. This model is known to exhibit anomalous scaling of the mean squared displacement (MSD) of the form $\sim t^{\frac{2}{2+\alpha}}$ and weak ergodicity breaking in the sense that ensemble averaged and time averaged MSDs do not converge. In this paper, we look at the extreme value statistics of this model and derive, for all $\alpha$, the exact probability distributions of the maximum spatial displacement $M(t)$ and arg-maximum $t_m(t)$ (i.e. the time at which this maximum is reached) till duration $t$. In the second part of our paper, we analyze the statistical properties of the residence time $t_r(t)$ and the last-passage time $t_{\ell}(t)$ and compute their distributions exactly for all values of $\alpha$. Our study unravels that the heterogeneous version $(\alpha \neq 0)$ displays many rich and contrasting features compared to that of the standard Brownian motion (BM). For example, while for BM $(\alpha =0)$, the distributions of $t_m(t),~t_r(t)$ and $t_{\ell}(t)$ are all identical (\textit{\'{a} la} ``arcsine laws" due to L\'{e}vy), they turn out to be significantly different for non-zero $\alpha$. Another interesting property of $t_r(t)$ is the existence of a critical $\alpha$ (which we denote by $\alpha _c=-0.3182$) such that the distribution exhibits a local maximum at $t_r = t/2$ for $\alpha < \alpha _c$ whereas it has minima at $t_r = t/2$ for $\alpha \geq \alpha _c$. The underlying reasoning for this difference hints at the very contrasting natures of the process for $\alpha \geq \alpha _c $ and $\alpha < \alpha _c $ which we thoroughly examine in our paper. All our analytical results are backed by extensive numerical simulations.   
\end{abstract}

\begin{document}

\title{Extreme value statistics and arcsine laws for heterogeneous diffusion processes}
\author{Prashant Singh}
\email{prashant.singh@icts.res.in}
\affiliation{International Centre for Theoretical Sciences, Tata Institute of Fundamental Research, Bengaluru 560089, India}

\date{\today}

\maketitle
\section{INTRODUCTION}
Many real-world systems involve motion of tracer particles in a heterogeneous medium with substantial spatial variations of the diffusion coefficient. For example, in \bluew{\cite{Kuhn2011, English2011}}, the dynamics of protein in the cell cytoplasm was shown to exhibit a systematic spatial variation of the diffusion coefficient by using mesoscopic numerical methods. Similarly, the mobility of cajal bodies (nuclear organelles) inside living cells develops heterogeneity due to their interactions with other nuclear components \bluew{\cite{Platani2002}}. Other examples of heterogeneous diffusion include particle moving between nearly parallel plates \bluew{\cite{Lancon2001}}, diffusion in presence of temperature gradient \bluew{\cite{Yanga2012}}, diffusion in nanoporous solids \bluew{\cite{Maris2021}}, confined hard sphere fluid \bluew{\cite{Mittal2008}} and so on. Descriptions with space-dependent diffusion coefficient have also been useful in modelling the diffusion in turbulent media \bluew{\cite{Richardson1926}} and on fractal objects \bluew{\cite{Shaughnessy1985}}. Quite recently, several studies on heterogeneous diffusive processes (HDPs) have revealed anomalous scaling of the mean squared displacement (MSD) and weak ergodicity breaking between time averaged and ensemble averaged MSDs \bluew{\cite{Cherstvy2013, Cherstvy2014, Leibovich2019,Cherstvy2015, Wang2019, Fa2005, new2}}. Persistent properties of HDPs have also been investigated in \bluew{\cite{Singh2020}}. Extensions of HDPs driven by colored noises were considered in \cite{MutothyaXu2021,Mutothya2021,Xu2020}. \greenw{Rigorous efforts have also been made to understand the combined effect of HDP and other models like fractional Brownian motion \cite{new1} and scaled Brownian motion \cite{Rnew1}}.

\begin{figure}[t]
\includegraphics[scale=0.34]{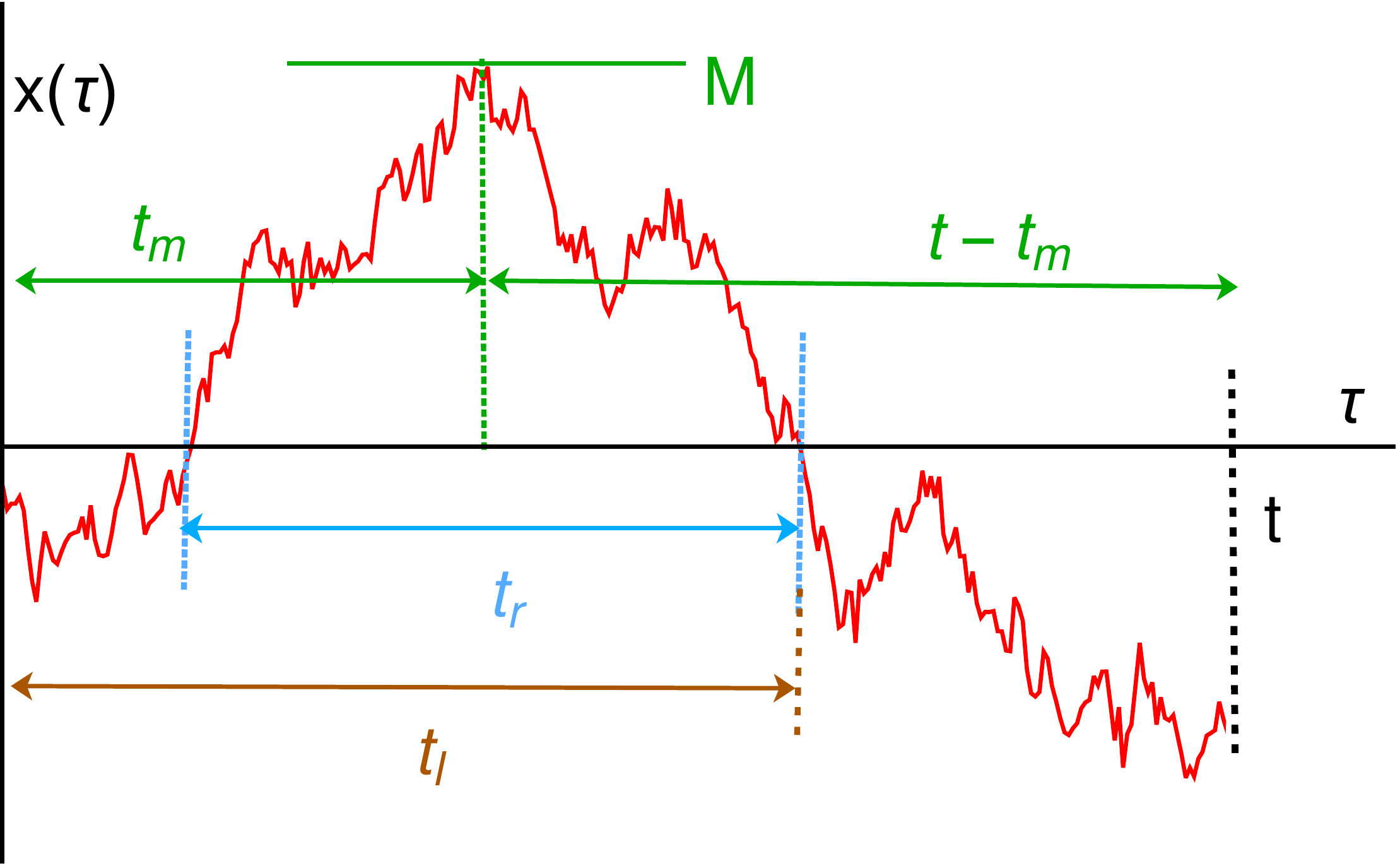}
\centering
\caption{An schematic illustration of the maximum distance $M$ attained by the process $x(\tau)$ (shown in red) in Eq. \eqref{model-eq-1} till duration $t$, i.e. $M(t) = \text{max}[\{ x (\tau) \}],$ where $0 \leq  \tau \leq t$. The time $t_m$ represents the time at which this maximum is attained. The time $t_r$ represents the amount of time for which $x(\tau)$ stays in the positive semi-axis and the duration $t_{\ell}$ is the last time that the process changes its sign (or crosses the origin).}
\label{trajectory-pic-1}
\end{figure}

In this paper, we analyse a simple model of one dimensional HDP where the diffusion coefficient has a power-law dependence on the position of the particle, i.e. $D(x) \sim |x|^{-\alpha}$ with $\alpha >-1$. For $\alpha =0$, it reduces to the homogeneous case of standard Brownian motion (BM). While the BM is extensively studied in the literature and a large number of results are known, the amount of studies for HDPs is still far from exhaustive. In an attempt towards this direction, we here investigate the extreme value statistics (EVS) of the HDP with power-law form for $D(x)$. In particular, we study how heterogeneity ramifies the statistics of the maximum $M(t)$ of the trajectory $x(t)$ observed till time $t$ and the time $t_m(t)$ at which this maximum is achieved. A schematic illustration of $M(t)$ and $t_m(t)$ for a trajectory is shown in Figure \ref{trajectory-pic-1}.

For one dimensional BM $(\alpha = 0)$, the marginal distributions of $M(t)$ and $t_m(t)$ read \bluew{\cite{Levy}}
\begin{align}
& P_m(M|t) = \frac{1}{\sqrt{\pi D_0 t}}~\text{exp} \left( -\frac{M^2}{4 D_0 t}\right) ,
\label{extreme-eq-20-ne-1} \\
& \mathcal{P}_m(t_m|t) = \frac{1}{\pi \sqrt{t_m(t-t_m)}},\label{extreme-eq-20-ne-2}
\end{align}
where $D_0$ is the diffusion coefficient. Beyond BM, such studies have also been performed for other stochastic processes like constraint Brownian motion, random walk and their generalisations \bluew{\cite{Levy,tmax-2, EVS-correlated-2,EVS-correlated-3, EVS-correlated-4, Andersen, Rambeau11, PrashantArnab2021, MoriMajmax2021}}, random acceleration \bluew{\cite{tmax-RAP, Burkhardt1993, Singhrandom2020}}, active particles \bluew{\cite{tmax-RTP-1, Mori2020}}, fractional Brownian motion \bluew{\cite{tmax-FBM-1, tmax-FBM-3, tmax-anamolous}}, continuous time random walk \bluew{\cite{tmax-CTRW}}, random matrices \bluew{\cite{RM-1, RM-2, RM-3}}, fluctuating interfaces \bluew{\cite{KPZ-1, KPZ-2, tmax-interface-growth}}, transport models \bluew{\cite{EVS-con-1, EVS-STR, EVS-con-4}}, finance \bluew{\cite{tmax-3}} and other physical systems \bluew{\cite{disorder-1, spin,EVS-con-3,EVS-con-5, EVS-con-6}} (see \bluew{\cite{EVS-review-1, EVS-review-2, EVS-review-3, EVS-review-4, EVS-review-5, EVS-review-6, EVS-review-7, EVS-review-cor, Sabhapanditrev, Revsatya}} for review). The subject of extreme value statistics has found applications in ecology \bluew{\cite{ecology}}, computer science \bluew{\cite{tree, CS-1, CS-2}} and convex hull problems \bluew{\cite{Furling2009}}. Generalising these studies, the statistics of the time between maximum and minimum spatial displacements was also recently considered for Brownian motion and random walks in \bluew{\cite{mori2019, MoriMaj2020}}.  

Even though there has been a substantial amount of study on EVS, most of these studies, however, are based on homogeneous setup. On the other hand, we saw above that, in many physical situations, the heterogeneous description becomes more relevant. A natural question then is - what happens to the distributions of $M(t)$ and $t_m(t)$ in Eqs. \eqref{extreme-eq-20-ne-1} and \eqref{extreme-eq-20-ne-2} when the dynamics takes place in a heterogeneous medium? Our work aims to provide a systematic understanding to this question in the context of HDPs with power-law form of $D(x)$. Our study demonstrates that the extremal statistics of this model is rather rich and possesses many contrasting features compared to that of the BM.

In the second part of our paper, we investigate the properties of the following two quantities measured along a trajectory $x(t)$ observed till time $t$: (i) residence time $t_r(t)$ spent on the positive (or negative) semi axis and (ii) last time $t_{\ell}(t)$ that the particle crosses the origin. For a trajectory of the particle, these two quantities are illustrated in Figure \ref{trajectory-pic-1}. The celebrated \textit{arcsine laws} for one dimensional Brownian motion state that the probability distributions of $t_m(t)$, $t_r(t)$ and $t_{\ell}(t)$ are all exactly same and given by \bluew{\cite{Levy, Majumdar005}}
\begin{align}
 \mathcal{P}_i(t_i|t) = \frac{1}{\pi \sqrt{t_i(t-t_i)}},
 \label{eq-1}
\end{align}
where $t_i \in \{ t_m, t_r, t_{\ell}  \}$. The corresponding cumulative probability has the `arcsine' form
\begin{align}
\text{Prob}[t_i \leq t] = \frac{2}{\pi} \text{arcsine}\left( \sqrt{\frac{t_i}{t}}\right),
\label{eq-2}
\end{align}
and hence the name \textit{arcsine laws}. Over the years, these quantities have been studied in different contexts like Brownian motion, random walks and their generalisations \bluew{\cite{Comtetlast2020, Majumdar005, Dhar99,Majumdar02, SMajumdar02, Sabhapandit06, Hollander2019, Comtet2003}}, random acceleration \bluew{\cite{tmax-RAP, Boutcheng16}}, continuous time random walk \bluew{\cite{Carmi2010, tmax-CTRW}}, fractional Brownian motion \bluew{\cite{tmax-FBM-3}}, run and tumble particle \bluew{\cite{Bressloff2020, tmax-RTP-1}}, finance \bluew{\cite{Charles80, Shiryaev02, tmax-3}}, renewal processes and other processes \bluew{\cite{Baldassarri99, Godreche01,Burov11,Kasahara77, Lamperti58}}. Quite recently, \textit{arcsine laws} have also been studied both experimentally and theoretically in stochastic thermodynamics \bluew{\cite{Barato18,dey2021}}. The statistics of residence time has also been used to classify the non-ergodicity in continuous-time random walk models \bluew{\cite{Bel2005, Barkai2006}}. Here, we look at the statistics of $t_r(t)$ and $t_{\ell}(t)$ in conjunction with $t_m(t)$ for the HDPs with power-law form of $D(x)$. More specifically, our interest is to study the ramifications of heterogeneity on the distributions of these three observables. We find that while their distributions are exactly same for $\alpha =0$ (BM), they turn out to be significantly different for non-zero $\alpha$. Our work provides the exact expression of the probability distributions of $M(t)$, $t_m(t)$, $t_r(t)$ and $t_{\ell}(t)$ for all $\alpha >-1$.

The remainder of paper is structured as follows: We define our model in Sec. \ref{model} and also present all our main results here. Sec. \ref{extreme-statistics} presents derivation of the joint distribution of $M(t)$ and $t_m(t)$ which is then used to obtain the marginal distribution of $M(t)$ in Sec. \ref{MargM} and that of $t_m(t)$ in Sec. \ref{MargtM}. We next compute the distributions of residence time $t_r(t)$ and last-passage time $t_{\ell}(t)$ in Secs. \ref{residence-time} and \ref{last-passage-time} respectively. Finally, we conclude in Sec. \ref{conclusion}.

\section{Model and summary of results}
\label{model}
We study the motion of a particle in one dimension moving in a heterogeneous medium. The heterogeneity is administered by considering the position-dependent diffusion coefficient $D(x)$. The time evolution equation for the position of the particle reads
\begin{align}
\frac{dx}{dt} = \sqrt{2 D(x)} ~\eta(t),
\label{model-eq-1}
\end{align}
where $\eta(t)$ is the Gaussian white noise with zero mean and correlation $\langle \eta(t) \eta(t') \rangle = \delta(t-t')$. In this paper, we focus on the power-law form of the diffusion coefficient:
\begin{align}
D(x) =  \frac{D_0~l^{\alpha}}{|x|^{\alpha}},~~~~\text{with }\alpha >-1,
\label{model-eq-2}
\end{align}
where $D_0$ is a positive constant that sets the strength of the noise and $l$ is the length scale over which $D(x)$ changes. The exponent $\alpha$ quantifies the strength of the gradient of $D(x)$. Throughout this paper, we consider $\alpha >-1$ and choose the starting position at the origin (unless specified). Note that for $\alpha =0$, $D(x)=D_0$ is just a constant and we recover the standard Brownian motion (BM). However, our main interest, in this paper, lies in $\alpha \neq 0$ and compare it with the BM.

Recall that the Langevin equation \eqref{model-eq-1} does not uniquely specify the model for position-dependent diffusion coefficient and one also needs to specify the sense in which the stochastic integration of Eq. \eqref{model-eq-1} is carried out \bluew{\cite{Vankampen1981, Lau2007}}. Throughout this paper, we will interprete Eq. \eqref{model-eq-1} in Ito sense. \greenw{ Another problem that one encounters in simulation is $D(x)$ diverges as $|x| \to 0$ for $\alpha >0$ while it tends to zero for $\alpha <0$. This will cause the shooting off of particles to the infinity for $\alpha >0$ or accumulation around $x=0$ for $\alpha < 0$.} In order to avoid this problem in simulation, we deploy the following form of $D(x)$ in simulation:
\begin{align}
D(x) = \lim _{x_{\text{off}} \to 0^+} \frac{D_0 l^{\alpha}}{\left(|x| + x_{\text{off}} \right)^{\alpha}},~~~\text{(for simulation)}.
\label{model-eq-3}
\end{align}
On the other hand, for all our analytic calculations, we take the form of $D(x)$ in Eq. \eqref{model-eq-2}. Such $x_{\text{off}}$-considerations of $D(x)$ with power-law form have also been studied in \bluew{\cite{Cherstvy2013,Cherstvy2014, Leibovich2019}}. For general $\alpha$, the mean-squared displacement of $x(t)$ in Eq. \eqref{model-eq-1} scales with time as $\langle x^2(t) \rangle \sim t^{\frac{2}{2+\alpha}}$ implying sub-diffusive behaviour for $\alpha >0$, super-diffusive for $\alpha <0$ and diffusive for $\alpha =0$ \bluew{\cite{Cherstvy2014}}. Recently, this model was shown to display weak ergodicity breaking in the sense that time averaged and ensemble averaged MSDs are not identical \bluew{\cite{Cherstvy2013}}.

Here, we look at the statistical properties of the maximum value $M(t)$ that the position $x(t)$ of the particle attains till duration $t$ i.e., $M(t) = \text{max}[\{ x (\tau) \}],$ where $0 \leq  \tau \leq t$ [see Figure \ref{trajectory-pic-1}]. In conjunction to this, we also investigate the statistics of the time $t_m(t)$ at which this maximum is reached. Exploiting the path-decomposition method for Markov processes \bluew{\cite{tmax-2}}, we derive exact expression for the joint distribution of $M(t)$ and $t_m(t)$ for all values of $\alpha$. Marginalising this joint distribution provides the exact form of the distributions of $M(t)$ and $t_m(t)$.  

Next, we also look at the statistics of residence time $t_r(t)$ which refers to the amount of time that the particle stays in $x>0$ region till duration $t$. Formally, it is written as $t_r(t) = \int _{0}^{t} d\tau~ \Theta \left( x(\tau) \right)$, where $\Theta (x)$ denotes the Heaviside theta function. Using the Feynman-Kac formalism \bluew{\cite{Kac1,Majumdar005}}, we compute the exact probability distribution of $t_r(t)$ for all values of $\alpha$. Finally, we study the last time $t _{\ell}(t)$ that the process $x(\tau)$ in Eq. \eqref{model-eq-1} changes sign (or crosses the origin) till duration $t$ and derive exact probability distribution for  $t _{\ell}(t)$. An schematic illustration of $M,~t_m,~t_r$ and $t_{\ell}$ for a typical trajectory of the particle is shown in Figure \ref{trajectory-pic-1}.  Below, we summarise our main results:
\begin{figure}[t]
\includegraphics[scale=0.37]{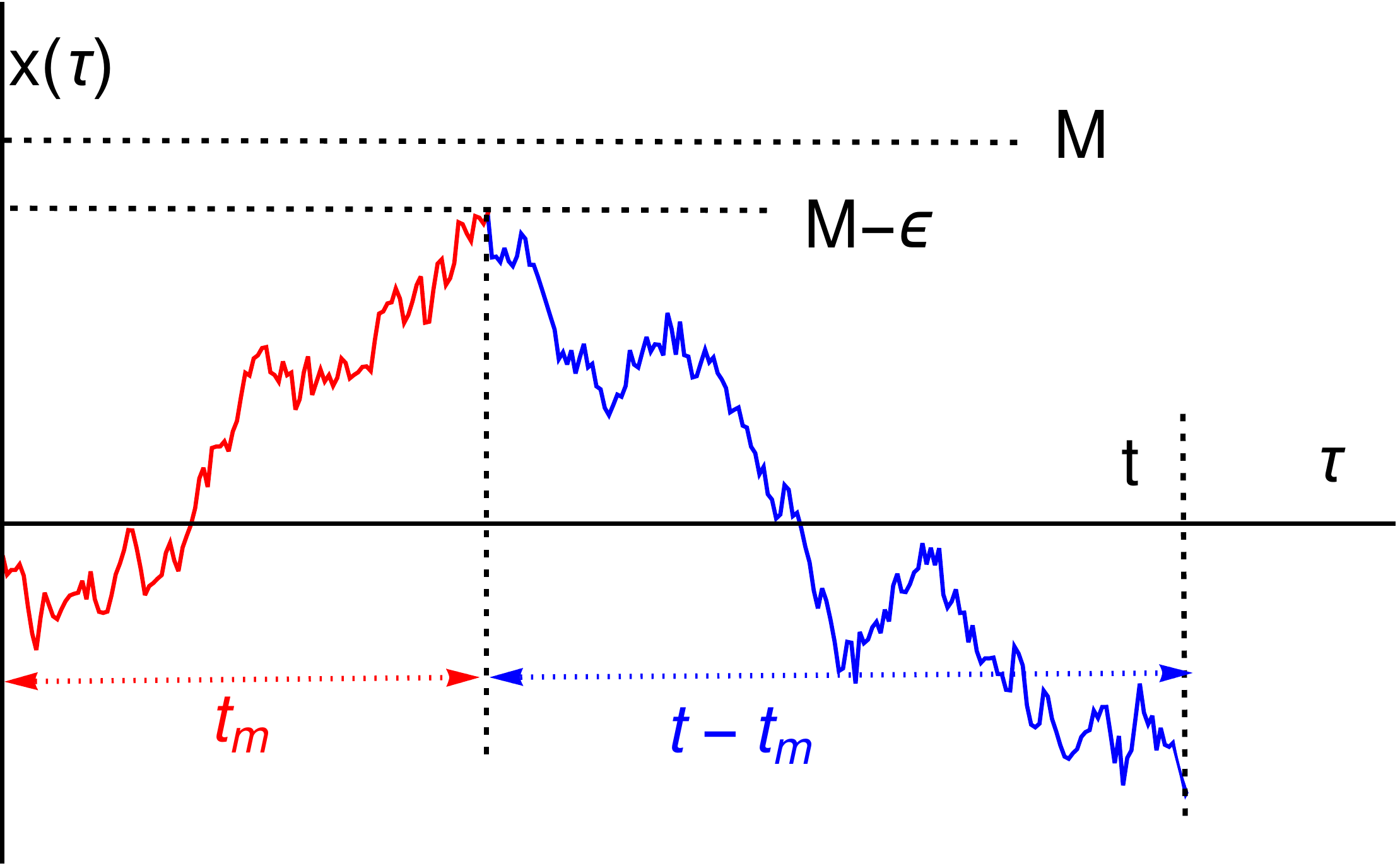}
\centering
\caption{Schematic of a typical trajectory of the particle in which it reaches $M-\epsilon$ at time $t_m$ for the first time and remains below $M$ in the remaining time $t-t_m$. The trajectory can be decomposed into two parts: from $0$ to $t_m$ (shown in red) and from $t_m$ to $t$ (shown in blue).}
\label{trajectory-pic-2}
\end{figure}

\begin{enumerate}
\item For general $\alpha$, we derive the exact probability distribution of the maximum $M$. Denoting this distribution by $P_m(M|t)$, we show that it possesses a scaling behaviour of the form
\begin{align}
P_m(M|t) = \frac{1}{\left( \mathcal{D}_{\alpha} t\right)^{\frac{1}{2+\alpha}}} \mathcal{F}_{\alpha} \left( \frac{M}{\left( \mathcal{D}_{\alpha} t\right)^{\frac{1}{2+\alpha}}} \right),
\label{extreme-eq-17} 
\end{align}
where $\mathcal{D}_{\alpha}$ is a constant given in Eq. \eqref{extreme-eq-9} and the scaling function $\mathcal{F}_{\alpha}(z)$ is defined as
\begin{align}
~~~~~~~~~~& \mathcal{F}_{\alpha}(z) = \frac{\mathcal{H}_{\frac{1}{2+\alpha}} \left(0 \right)}{z^{3+\alpha}} \int _{0}^{\infty} dw ~e^{-\frac{w}{z^{2+\alpha}}} ~\mathbb{H}_{\frac{1}{2+\alpha}}(\sqrt{w}), \label{extreme-eq-18} 
\end{align}
with $ \mathcal{H}_{\frac{1}{2+\alpha}} \left(0 \right)=\frac{2^{\frac{1}{2+\alpha}}}{\Gamma \left( \frac{1+\alpha}{2+\alpha}\right)}$ and the function $\mathbb{H}_{\beta}(w)$ given in Eq. \eqref{appen-ILT-J-Eq-6}. For large $z$, we find that the scaling function decays as $\mathcal{F}_{\alpha}(z) \sim z^{\alpha} e^{-z^{2+\alpha}/4}$.

\item We next calculate the probability distribution of the time $t_m(t)$ and show that it possesses the scaling structure
\begin{align}
\mathcal{P}_m(t_m|t) = \frac{1}{t} ~\mathcal{G} _{m}^{\alpha} \left( \frac{t_m}{t}\right),
\label{extreme-res-eq-1}
\end{align}
with the scaling function $\mathcal{G} _{m}^{\alpha} \left( z \right)$ defined as
\begin{align}
~~~~~~~~\mathcal{G} _{m}^{\alpha} \left( z \right) = \frac{(2+\alpha)\mathcal{H}_{\frac{1}{2+\alpha}} \left(0 \right)}{2z^{\frac{1+\alpha}{2+\alpha}} (1-z)^{\frac{1}{2+\alpha}}}~\int _{0}^{\infty} dw\frac{\mathbb{X}_{\alpha} \left( \sqrt{\frac{1-z}{z w^{2+\alpha}}} \right)}{\mathcal{H}_{\frac{1}{2+\alpha}}\left( w^{\frac{2+\alpha}{2}}\right)}.
\label{extreme-res-eq-2}
\end{align}
The functions $\mathcal{H}_{\beta}(x)$ and $\mathbb{X}_{\beta}(x)$ are defined, respectively, in Eqs. \eqref{extreme-eq-8} and \eqref{appen-ILTYY-eq-7}. For BM, it follows from Eq. \eqref{extreme-eq-20-ne-2} that the scaling function $\mathcal{G} _{m}^{\alpha} \left( z \right)$ is symmetric under the transformation $z \to 1-z$. However, as evident from Eq. \eqref{extreme-res-eq-2}, this symmetry is no longer present for general $\alpha$. This is further exemplified by the behaviour of the scaling function as $z \to 0$ and $z \to 1$ for which we later show divergences of the form $\mathcal{G} _{m}^{\alpha} \left( z \to 0 \right) \sim z^{-\frac{1+\alpha}{2+\alpha}}$ and $\mathcal{G} _{m}^{\alpha} \left( z \to 1 \right) \sim (1-z)^{-\frac{1}{2}}$.

\item We also compute the distribution $\mathcal{P}_r(t_r|t)$ of the residence time $t_r$ for general $\alpha$ show that it has the scaling form
\begin{align}
\mathcal{P}_r(t_r|t) = \frac{1}{t} ~\mathcal{G} _{r}^{\alpha} \left( \frac{t_r}{t}\right),
\label{resi-eq-8}
\end{align}
where the scaling function $\mathcal{G} _{r}^{\alpha} \left( z \right)$ is given by
\begin{align}
\begin{split}
& ~~~~~~~~~~\mathcal{G} _{r}^{\alpha} \left( z \right) = \frac{ \sin \left( \frac{\pi}{2+\alpha}\right) }{\pi~ \left[ z(1-z) \right]^{\frac{1+\alpha}{2+\alpha}}} \\
&~~~~~~ \times \frac{1}{z^{\frac{2}{2+\alpha}}+(1-z)^{\frac{2}{2+\alpha}}+2  \cos \left( \frac{\pi}{2+\alpha}\right) \left[z(1-z) \right] ^{\frac{1}{2+\alpha}}}.
\end{split} 
\label{resi-eq-9}
\end{align}
This scaling function diverges as $\mathcal{G} _{r}^{\alpha} \left( z \right) \sim z^{-\frac{1+\alpha}{2+\alpha}}$ as $z \to 0$ and as $\mathcal{G} _{r}^{\alpha} \left( z \right) \sim (1-z)^{-\frac{1+\alpha}{2+\alpha}}$ as $z \to 1$.

\item Finally, we derive the probability distribution $\mathcal{P}_{ \ell } \left( t_{\ell}|t \right)$ of the last-passage time $t_{\ell}$ which also possesses the scaling structure
\begin{align}
\mathcal{P}_{ \ell } \left( t_{\ell}|t \right) = \frac{1}{t} \mathcal{G}_{\ell}^{\alpha} \left( \frac{t_{\ell}}{t}\right),
\label{last-passage-time-eq-7}
\end{align}
with the scaling function $\mathcal{G}_{\ell}^{\alpha} \left( z \right)$ given by
\begin{align}
\mathcal{G}_{\ell}^{\alpha} \left( z \right) = \frac{z^{-\frac{1+\alpha}{2+\alpha}} \left( 1-z \right) ^{-\frac{1}{2+\alpha}}}{\Gamma \left( \frac{1+\alpha}{2+\alpha}\right)~\Gamma \left( \frac{1}{2+\alpha}\right)}.
\label{last-passage-time-eq-9}
\end{align}
Here, once again, we find that the scaling function does not retain symmetry under the transformation $z \to 1-z$ for $\alpha \neq 0$. Consequently, we get different behaviours of $\mathcal{G}_{\ell}^{\alpha} \left( z \right)$ for $z \to 0$ and $z \to 1$, viz. $\mathcal{G} _{\ell }^{\alpha} \left( z \to 0 \right) \sim z^{-\frac{1+\alpha}{2+\alpha}}$ and $\mathcal{G} _{ \ell }^{\alpha} \left( z \to 1 \right) \sim (1-z)^{-\frac{1}{2+\alpha}}$.

\end{enumerate}
We remark that for $\alpha =0$, all scaling functions written above converge to that of the BM in Eqs. \eqref{extreme-eq-20-ne-1} and \eqref{eq-1}. Also, note that that the distributions of $t_m$, $t_r$ and $t_{\ell}$ are completely different for general $\alpha$ and they are identical only for $\alpha =0$. Another interesting property contrary to that of the BM is that the distributions of $t_m$ and $t_{\ell}$ for $\alpha \neq 0$ have asymmertic peaks (divergenecs) as $t_i \to 0^+$ and $t_i \to t^-$ where $t_i \in \{t_m, t_{\ell} \}$. All these observations exemplify that the properties of $M$, $t_m$, $t_r$ and $t_{\ell}$ for $\alpha \neq 0$ are remarkably different than that of the BM. In the following, we provide a detailed 
analysis of these quantities and point out the key differences for $\alpha \neq 0$.

\section{Extreme value $M$ and time $t_m$ to reach maximum}
\label{extreme-statistics}
Let us begin with the joint distribution $\mathcal{P}(M, t_m|t)$ of the maximum displacement $M$ and the time $t_m$ at which the maximum is attained till duration $t$. The initial position is fixed to the origin. To compute this distribution, we decompose the trajectory in two parts: (i) the part from $0$ to $t_m$ and (ii) the part from $t_m$ to $t$. They are shown schematically in Figure \ref{trajectory-pic-2}, where the red half corresponds to the part (i) and the blue half represents the part (ii). Since the process is Markovian, the two parts are statistically independent.

Let us now calculate the contribution of each part. In part (i), the particle reaches $x=M$ at time $t_m$ for the first time given that it was at the origin initially. Therefore, the probability weight in this part is just the first-passage time distribution $ F_{M}(t_m|0)$ to reach $M$ for the first time at $t_m$ given that the particle was initially at $x=0$. In part (ii), the process remains below $x=M$ in the interval $t-t_m$ such that it was at $x=M$ at time $t_m$. Hence, the weight of this part is given by the survival probability $S_{M}(t-t_m|M)$ where we have used the notation $S_{x_m}(\tau |x_0)$ to denote the probability that the particle has not crossed $x=x_m$ up to time $\tau$ starting from $x=x_0$. Note that the process remains below $x=M$ in both parts. As remarked before, these two contributions are statistically independent due to the Markovianity of the process. However, as shown later, it turns out that $S_{M}(\tau|M)=0$ for all non-zero $\tau$ which implies that the contribution from part (ii) is zero. To circumvent this problem, we follow the procedure in \bluew{\cite{tmax-2, tmax-RAP}} where we compute $F_{M-\epsilon}(t_m|0)$ and $ S_M(t-t_m|M-\epsilon)$ instead of $F_{M}(t_m|0)$ and $ S_M(t-t_m|M)$ and later take the $\epsilon \to 0^+$ limit. The joint distribution $P(M, t_m|t)$ can then be written as
\begin{align}
\mathcal{P}(M, t_m|t)=\frac{ F_{M-\epsilon}(t_m|0) ~S_M(t-t_m|M-\epsilon)}{\mathcal{N} (\epsilon)}.
\label{extreme-eq-1}
\end{align}
Here $1/\mathcal{N} (\epsilon)$ is the proportionality constant independent of $t$ and $t_m$ and fixed by the normalisation condition. For later calculations, it turns out useful to take double Laplace transformation of Eq. \eqref{extreme-eq-1} with respect to $t_m ~(\to p)$ and $t ~(\to s)$: 
\begin{align}
\bar{P}(M, p|s)= \frac{ \bar{F}_{M-\epsilon}(s+p|0) \bar{S}_M(s|M-\epsilon)}{\mathcal{N} (\epsilon)},
\label{extreme-eq-2}
\end{align}
where $\bar{P}(M, p|s)$ is the double Laplace transformation of $\mathcal{P}(M, t_m|t)$. Quite remarkably, using the Markovian property, we have completely specified the joint distribution of $M$ and $t_m$ in terms of its survival probability and first-passage time distribution. In what follows, we use the standard techniques to calculate these distributions and probabilities and then use Eq. \eqref{extreme-eq-2} to compute the joint distribution. 

\subsection{Survival probability $S_M(t|x_0)$}
Let us focus on the survival probability $S_M(t|x_0)$ for our model in Eq. \eqref{model-eq-1}. For simplicity, we consider $M \geq 0$ and $x_0 \leq M$ which is also consistent with our main aim of computing the joint distribution in Eq. \eqref{extreme-eq-2}. In Ito set-up, $S_M(t|x_0)$ obeys the backward Fokker Planck equation \bluew{\cite{Redner}}
\begin{align}
\partial _t S_M(t|x_0) = D(x_0) \partial _{x_0 } ^2 S_M(t|x_0),
\label{extreme-eq-3}
\end{align}
with $D(x_0)$ defined in Eq. \eqref{model-eq-2}. Our aim is to solve this equation for general $\alpha$. In order to solve this equation, we have to specify the appropriate initial condition and boundary conditions. Initially, the particle starts from the position $x_0$ which is different from the position of the absorbing wall at $x=M$.  Consequently, the particle always survives and we get
\begin{align}
S_M(0|x_0) = 1. \label{extreme-eq-4-new}
\end{align}
Next, we specify the boundary conditions which read 
\begin{align}
& S_M \left(t|x_0 \to M^- \right) = 0, \label{extreme-eq-4} \\
& S_M \left(t|x_0 \to -\infty \right) = 1. \label{extreme-eq-5}
\end{align} 
To understand the boundary condition in Eq. \eqref{extreme-eq-4}, note that if the particle initially starts from $x_0 \to M^-$, then it will immediately get absorbed. This results in the zero survival probability. On the other hand, if the particle is initally very far from the origin ($x_0 \to -\infty$), then it will survive the barrier at $x=M$ for all finite time. This gives rise to the second boundary condition in Eq. \eqref{extreme-eq-5}.
\greenw{Solving Eq. \eqref{extreme-eq-3} [see Sec. I of Supplementary material (SM) \cite{Supplementary} for details], we obtain the Laplace transformation of $S_M(t|x_0)$ for $x_0 \geq 0$ and $\alpha > -1$ as} 
\begin{align}
\bar{S}_M(s|x_0) = \frac{1}{s} \left[1- \frac{\mathcal{H}_{\frac{1}{2+\alpha}} \left( (a_s x_0)^{\frac{2+\alpha}{2}} \right)}{\mathcal{H}_{\frac{1}{2+\alpha}} \left( (a_s M)^{\frac{2+\alpha}{2}} \right)} \right],
\label{extreme-eq-7}
\end{align} 
where the functions $\mathcal{H}_{\beta} (x_0)$ and $a_s$ are defined as
\begin{align}
 & \mathcal{H}_{\beta} (x_0) = x_0 ^{\beta} \left[ I _{\beta}(x_0)+I_{-\beta} (x_0)\right],\label{extreme-eq-8} \\
 & a_s  = \left(\frac{s}{\mathcal{D}_{\alpha}} \right)^{\frac{1}{2+\alpha}},~~~\text{with }\mathcal{D}_{\alpha} = \frac{D_0 l^{\alpha} (2+\alpha)^2}{4}.\label{extreme-eq-9} 
\end{align}
Here, $I _{\beta}(x_0)$ is the modified Bessel function of first kind. In the next section, we proceed to use the Laplace transform $\bar{S}_M(s|x_0)$ from Eq. \eqref{extreme-eq-7} to calculate the joint distribution $\bar{P}(M, p|s)$ in Eq. \eqref{extreme-eq-2}.
\subsection{Joint probability distribution $\mathcal{P}(M, t_m|t)$}
Coming to the expression of $\bar{P}(M, p|s)$ in Eq. \eqref{extreme-eq-2}, we need to specify the Laplace transforms $\bar{S}_M(s|M-\epsilon)$ and $\bar{F}_{M-\epsilon}(s+p|0)$ in the limit $\epsilon \to 0^+$. Using Eq. \eqref{extreme-eq-7}, these Laplace transforms can be easily calculated. \greenw{We refer to Eqs. \eqref{extreme-eq-10} and \eqref{extreme-eq-11} in appendix  \ref{appen-exp} for the rigorous expression of these Laplace transforms.}
Plugging them in Eq. \eqref{extreme-eq-2} gives $\bar{P}(M, p|s)$ as
\begin{align}
\bar{P}(M, p|s)&= \frac{ \epsilon ~ \mathcal{H}_{\frac{1}{2+\alpha}} \left(0 \right) }{\mathcal{N} (\epsilon)s~\mathcal{H}_{\frac{1}{2+\alpha}} \left( (a_{s+p} M)^{\frac{2+\alpha}{2}} \right)}
\nonumber  \\
& ~~~~~~\times \frac{\partial _M \left[ \mathcal{H}_{\frac{1}{2+\alpha}} \left( (a_s M)^{\frac{2+\alpha}{2}} \right) \right]}{ \mathcal{H}_{\frac{1}{2+\alpha}} \left( (a_s M)^{\frac{2+\alpha}{2}} \right) } . \label{extreme-eq-12}
\end{align}
The task now is to evaluate the function $\mathcal{N}(\epsilon)$. For this, we use the normalisation condition of $\mathcal{P}(M,t_m|t)$ which in terms of the Laplace transform $\bar{P}(M, p|s)$ becomes
\begin{align}
\int _ 0^{\infty}~ \bar{P}(M, p=0|s)~dM = \frac{1}{s}. \label{extreme-eq-13}
\end{align}
Plugging $\bar{P}(M, p=0|s)$ from Eq. \eqref{extreme-eq-12}, it is easy to show that $\mathcal{N}(\epsilon)=\epsilon$. Substituting this in Eq. \eqref{extreme-eq-12} yields
\begin{align}
\bar{P}(M, p|s)&= \frac{  \mathcal{H}_{\frac{1}{2+\alpha}} \left(0 \right) }{s~\mathcal{H}_{\frac{1}{2+\alpha}} \left( (a_{s+p} M)^{\frac{2+\alpha}{2}} \right)}
\nonumber  \\
& ~~~~~~\times \frac{\partial _M \left[ \mathcal{H}_{\frac{1}{2+\alpha}} \left( (a_s M)^{\frac{2+\alpha}{2}} \right) \right]}{ \mathcal{H}_{\frac{1}{2+\alpha}} \left( (a_s M)^{\frac{2+\alpha}{2}} \right) } . \label{extreme-eq-14}
\end{align}
To summarise, we have exactly computed the Laplace transformation $\bar{P}(M, p|s)$ of the joint distribution of $M$ and $t_m$ for all values of $\alpha$. To get distribution in the time domain, one has to perform double inverse Laplace transformations which, unfortunately, turns out to be challenging. However, one could still obtain the explicit expressions of the marginal distributions of $M(t)$ and $t_m(t)$ by appropriately integrating $\bar{P}(M, p|s)$ in Eq. \eqref{extreme-eq-14}. In what follows, we use $\bar{P}(M, p|s)$ to obtain the marginal distribution of the maximum $M(t)$ followed by that of the arg-maximum $t_m(t)$. 
\begin{figure*}[t]
\includegraphics[scale=0.3]{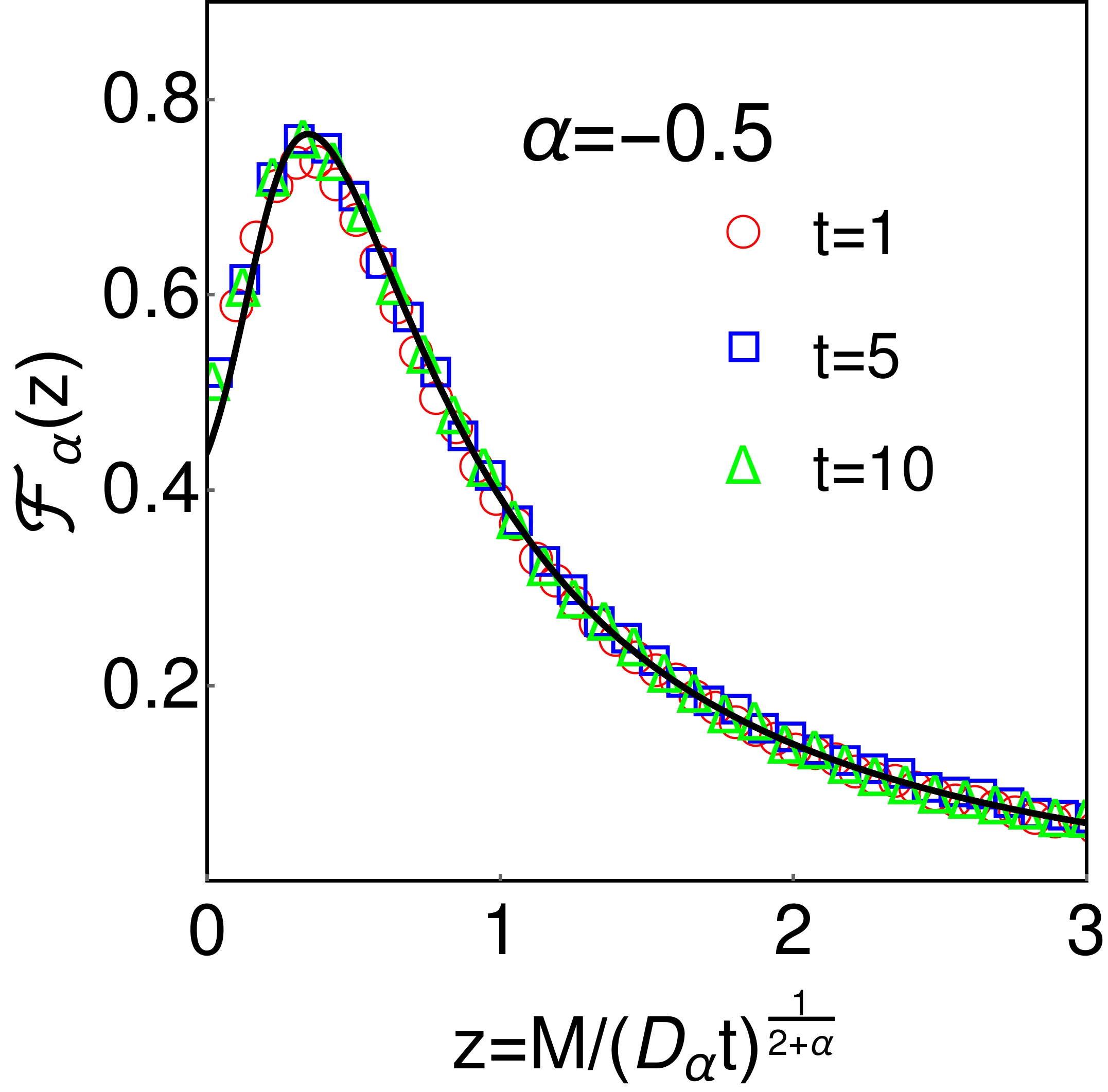}
\includegraphics[scale=0.3]{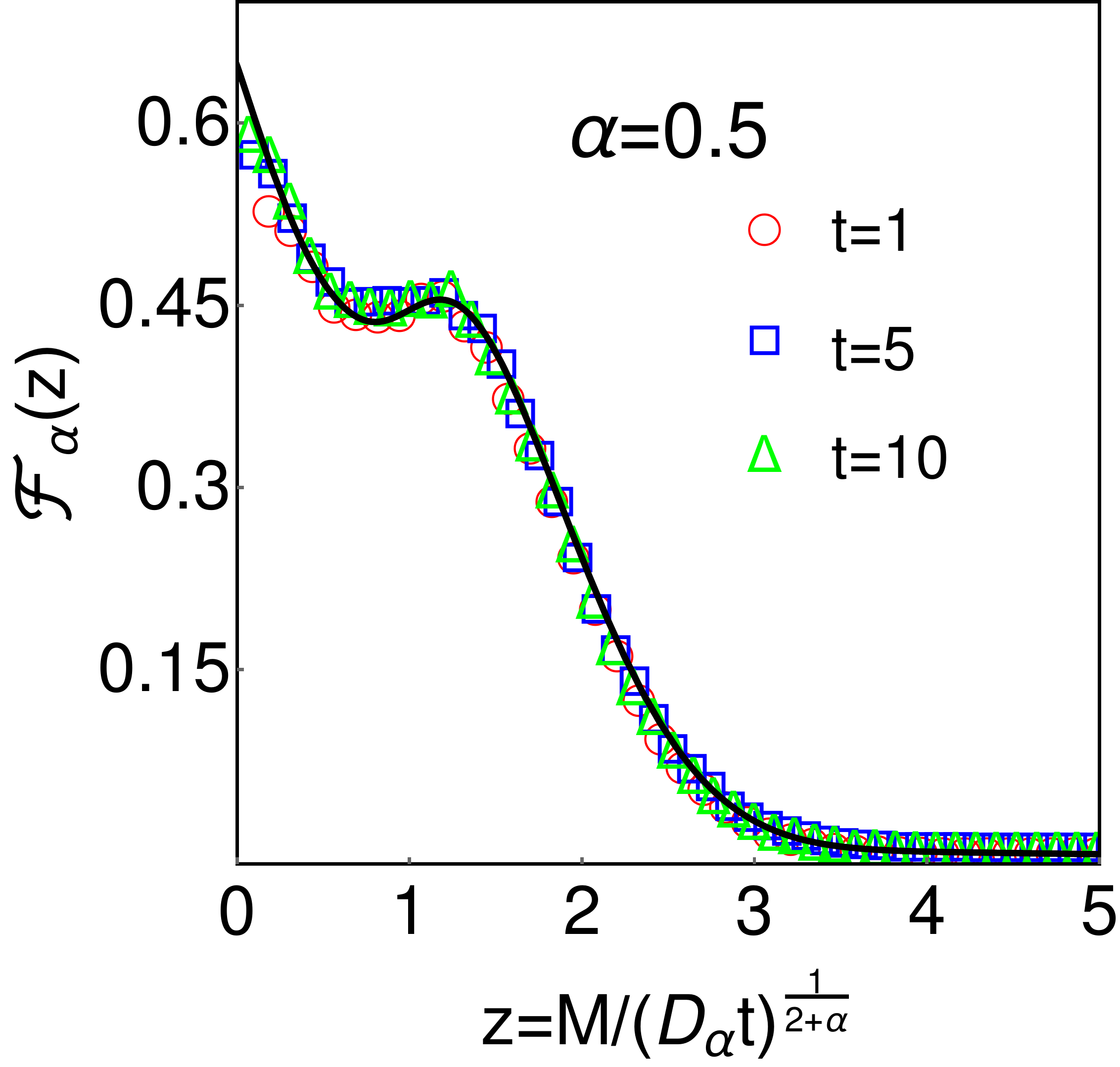}
\centering
\caption{Scaling function $\mathcal{F}_{\alpha}(z)$ in Eq. \eqref{extreme-eq-18} is plotted for two values of $\alpha$. In both panels, solid black line represents the analytic expression in Eq. \eqref{extreme-eq-18} and the symbols represent simulation data. We have chosen $D_0=0.1$ and $l=1$ }
\label{scaling-M-fig}
\end{figure*}
\subsection{Marginal distribution $P_m(M|t)$ of $M(t)$}
\label{MargM}
To get the marginal distribution  $P_m(M|t)$ of the maximum $M$, we integrate the joint distribution $ \mathcal{P}(M,t_m|t)$ over all $t_m$. In terms of the Laplace variables $p$ and $s$, this is equivalent to putting $p=0$ in the expression of $\bar{P}(M, p|s)$ in Eq. \eqref{extreme-eq-14} which then gives the Laplace transformation $\bar{P}_m(M|s)$ of the distribution $P_m(M|t)$. One finds
\begin{align}
\bar{P}_m(M|s)& = \bar{P}(M, p=0|s), \nonumber \\
& =-\frac{d \bar{J}(M,s)}{dM},~~~~\text{with} \label{extreme-eq-15} \\
\bar{J}(M,s)&= \frac{\mathcal{H}_{\frac{1}{2+\alpha}} \left(0 \right)}{s ~\mathcal{H}_{\frac{1}{2+\alpha}} \left( (a_s M)^{\frac{2+\alpha}{2}} \right)}, \label{extreme-eq-16} 
\end{align}
where the function $\mathcal{H}_{\beta}(x_0)$ in the last equation is defined in Eq. \eqref{extreme-eq-8}. We now proceed to perform the inverse Laplace transformation of $\bar{P}_m(M|s)$ in Eq. \eqref{extreme-eq-15}. Fortunately, this inversion can be exactly carried out. We refer to Sec. II of SM \cite{Supplementary} for the details of this calculation. The distribution $P_m(t_m|t)$ possesses the scaling structure as written in \eqref{extreme-eq-17} with the scaling function $\mathcal{F}_{\alpha}(z)$ defined in \eqref{extreme-eq-18}.

Few remarks are in order. Firstly, for $\alpha =0$ in Eq. \eqref{extreme-eq-18}, we find $\mathbb{H}_{\frac{1}{2}}(w) $ in Eq. \eqref{appen-ILT-J-Eq-6} has the simple form 
\begin{align}
\mathbb{H}_{\frac{1}{2}}(w) = \sqrt{\frac{2}{\pi}} \sin (w).
\end{align}
Plugging this in the expression of the scaling function $\mathcal{F}_{\alpha}(z)$ in \eqref{extreme-eq-18} and performing the integration over $w$ yields
\begin{align}
\mathcal{F}_{\alpha}(z) = \frac{e^{-z^2/4}}{\sqrt{\pi}} ,~~~~~(\text{for }\alpha =0).
\label{extreme-eq-20}
\end{align}
This matches with the distribution of $M$ for the standard Brownian motion in Eq. \eqref{extreme-eq-20-ne-1}. However, for general $\alpha$, the scaling function is given in Eq. \eqref{extreme-eq-18}. \greenw{We also remark that the scaling of the maximum with time as $M \sim t^{\frac{1}{2+\alpha}}$ in Eq. \eqref{extreme-eq-17} is quite expected since the position scales as $x \sim t^{\frac{1}{2+\alpha}}$. However, our analysis goes beyond this scaling behaviour and also provides an exact form of the associated scaling function for all values of $\alpha$.}

In Figure \ref{scaling-M-fig}, we have plotted $\mathcal{F}_{\alpha}(z)$ and compared it against the simulation for two different values of $\alpha$. For each value, we have conducted simulation for three different values of $t$. We see excellent match of our analytic result in Eq. \eqref{extreme-eq-18} with the simulations. To contrast these results with that of the standard Brownian motion, we look at the asymptotic behaviour of $\mathcal{F}_{\alpha}(z)$ for different $z$. In particular, for $\alpha \neq 0$, we have shown in Sec. III of SM \cite{Supplementary} that the scaling function has the following asymptotic forms:
\begin{align}
\mathcal{F}_{\alpha}(z) &\simeq \frac{1}{C_{\alpha} \Gamma \left( \frac{1+\alpha}{2+\alpha}\right)}-\frac{2z}{C_{\alpha} ^2 \Gamma\left( \frac{\alpha}{2+\alpha}\right)},~~\text{as } z \to 0, \label{extreme-eq-21} \\
&\simeq \frac{(2+\alpha)\mathcal{H}_{\frac{1}{2+\alpha}}(0) }{2^{\frac{3+2\alpha}{2+\alpha}}}  z^{\alpha} ~e^{-\frac{z^{2+\alpha}}{4} },~~\text{as } z \to \infty,
\label{extreme-eq-22}
\end{align}
where $C_{\alpha} = \frac{2^{2/2+\alpha}}{(2+\alpha)} \frac{\Gamma \left( \frac{1}{2+\alpha}\right)}{\Gamma \left( \frac{1+\alpha}{2+\alpha}\right)}$. On the other hand for $\alpha =0$, one gets from Eq. \eqref{extreme-eq-20}
\begin{align}
\mathcal{F}_{0}(z) &\simeq \frac{1}{\sqrt{\pi}} \left( 1-\frac{z^2}{4}\right), ~~~~\text{as } z \to 0, \label{extreme-eq-22-new}\\
& = \frac{e^{-z^2/4}}{\sqrt{\pi}},~~~~~~~~~~~~~~\text{as } z \to \infty. \label{extreme-eq-23-new}
\end{align}
We see that while, for $\alpha =0$, the scaling function decreases quadratically with $z$ as $z \to 0$ , it changes linearly for $\alpha \neq 0$ [see Eq. \eqref{extreme-eq-21}]. Also, the small $z$ behaviour is rather different for $\alpha <0$ and $\alpha >0$. For $\alpha <0$, we see, in Figure \ref{scaling-M-fig} (left panel), that $\mathcal{F}_{\alpha}(z)$ rises initially with $z$, attains a maximum value and then decreases again for large $z$. On the other hand, for $\alpha >0$, we see that $\mathcal{F}_{\alpha}(z)$ initially decreases with $z$, then rises at some intermediate $z$ until it attains a local maximum. After that, it again decreases for large $z$ (see Figure \ref{scaling-M-fig} (right panel)). Quite interestingly, we see a non-monotonic dependence of $\mathcal{F}_{\alpha}(z)$ on $z$ for all $\alpha \neq 0$ (see Figure \ref{scaling-M-fig}). However, as illustrated in Eq. \eqref{extreme-eq-20}, the scaling function decreases monotonically with $z$ for the BM for all values of $z$.
\begin{figure}[t]
\includegraphics[scale=0.35]{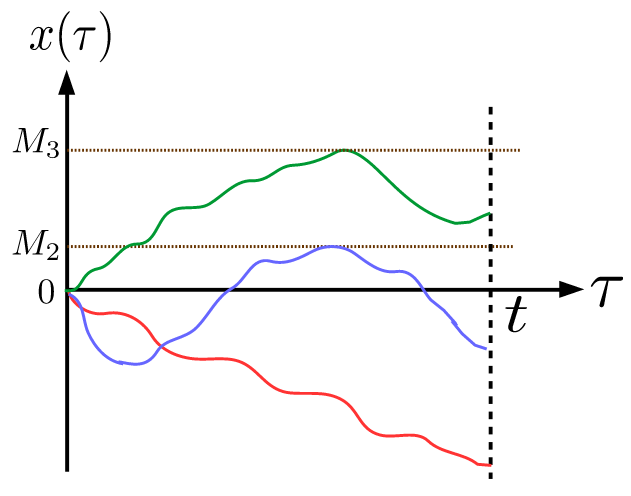}
\centering
\caption{Schematic of the trajectories that give rise to various values of the maximum $M$. For red trajectory, the particle stays below origin for all $t$. Consequently $M=0$. Similarly, for blue and green trajectories, the values of maximum are $M_1 ~(\ll t^{\frac{1}{2+\alpha}})$ and $M_2 ~(\gg t^{\frac{1}{2+\alpha}})$ respectively.  }
\label{non-mono-scal-arg-fig}
\end{figure}

\greenw{To understand the non-monotonic nature of the scaling function $\mathcal{F}_{\alpha}(z)$, let us analyse the trajectories that give rise to different values of the maximum $M(t)$. For simplicity, we focus on $\alpha <0$. In Figure \ref{non-mono-scal-arg-fig}, we have shown a schematic illustration of three coloured trajectories which contribute to three different maxima. The red trajectory stays below origin for all time which contributes to $M=0$. On the other hand, the blue trajectory contributes non-zero $M_1 ~(\ll t^{\frac{1}{2+\alpha}})$. Since for $\alpha <0$, the particle typically stays near the origin, it is more likely that it crosses the origin some number of times. Consequently, the likelihood of finding a red trajectory (where particle does not cross the origin) is less compared to a blue trajectory (where it crosses the origin a few times). In terms of the maximum $M$, this amounts to a smaller value of the distribution $P_m(M|t)$ for $M=0$ as compared to the non-zero $M$ [see Figure \ref{scaling-M-fig} (left panel)]. However, to obtain large values of $M$, the fluctuations have to be sufficiently strong to take it far away from the origin on the positive side (see green trajectory in Figure \ref{non-mono-scal-arg-fig}). Such fluctuations for $\alpha <0$ is extremely rare which results in smaller values of the distribution $P_m(M|t)$ for large $M$. Overall, we obtain a non-monotonic nature of $P_m(M|t)$ (or equivalently $\mathcal{F}_{\alpha}(z)$). Although we have presented the physical reasoning for $\alpha <0$, it is easy to extend it for $\alpha >0$ also.}    

\subsection{Marginal distribution $\mathcal{P}_m \left( t_m|t \right)$ of $t_m(t)$}
\label{MargtM}
This section deals with the probability distribution $\mathcal{P}_m \left( t_m|t \right)$ of the arg-maximum $t_m$. Let us denote its double Laplace transformation by $\bar{\mathcal{P}}_m(p|s)$. Marginalising $\bar{P}(M, p|s)$ in Eq. \eqref{extreme-eq-14} by integrating over all $M$, we find
\begin{align}
\bar{\mathcal{P}}_m(p|s) &= \int _{0}^{\infty} dM \frac{  \mathcal{H}_{\frac{1}{2+\alpha}} \left(0 \right) }{s~\mathcal{H}_{\frac{1}{2+\alpha}} \left( (a_{s+p} M)^{\frac{2+\alpha}{2}} \right)}
\nonumber  \\
& ~~~~~~\times \frac{\partial _M \left[ \mathcal{H}_{\frac{1}{2+\alpha}} \left( (a_s M)^{\frac{2+\alpha}{2}} \right) \right]}{ \mathcal{H}_{\frac{1}{2+\alpha}} \left( (a_s M)^{\frac{2+\alpha}{2}} \right)}.
\label{extreme-eq-30}
\end{align}
Recall that the function $\mathcal{H}_{\beta}(x_0)$ is defined in Eq. \eqref{extreme-eq-8}.
\greenw{To get the distribution in the time domain, one then needs to perform the double inverse Laplace transformation of Eq. \eqref{extreme-eq-30}. In Sec. IV of SM \cite{Supplementary}, we have explicitly carried out this inversion.} The final form of the distribution shows that $\mathcal{P}_m \left( t_m|t \right)$ possesses the scaling form as quoted in Eq.\eqref{extreme-res-eq-1} where the scaling function $\mathcal{G}_m^{\alpha}(z)$ is given in Eq. \eqref{extreme-res-eq-2}.

\begin{figure}[t]
\includegraphics[scale=0.35]{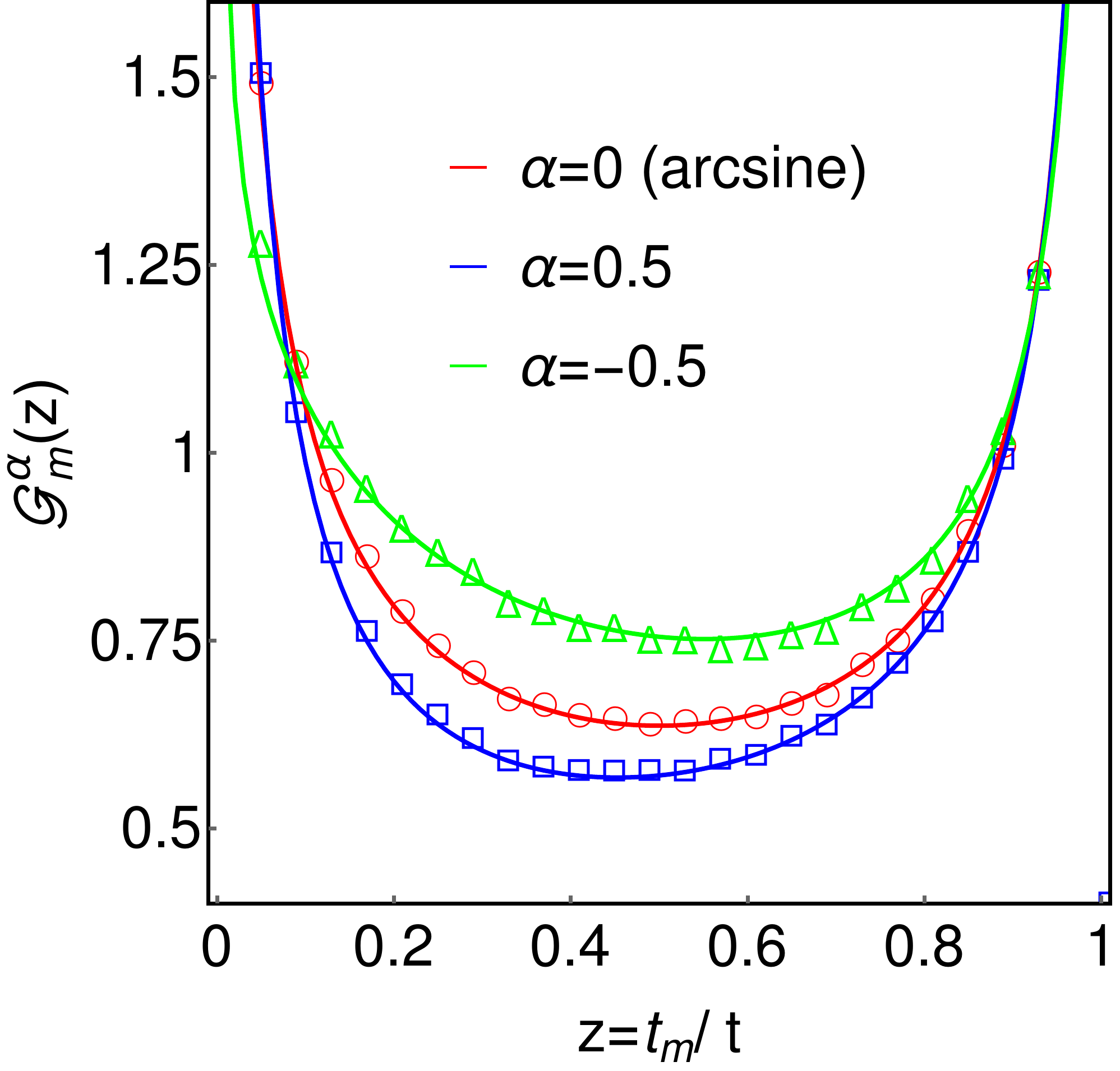}
\centering
\caption{We have plotted the scaling function $\mathcal{G}_m^{\alpha}(z)$ of arg-maximum $t_m$ for three different values of $\alpha$. The symbols represent the simulation data for $t=5$ which are compared with the analytic expression (shown by solid line) in Eq. \eqref{extreme-res-eq-2}. Parameters chosen are $D_0 = 0.1$ and $l=1$.}
\label{timemax-fig}
\end{figure}

In Figure \ref{timemax-fig}, we have illustrated this scaling behaviour for different values of $\alpha$. For all values, we see excellent agreement of our analytic results with the numerical simulations. To recover \textit{arcsine law} for $\alpha =0$, we notice that $\mathbb{X}_{\alpha}(x) = \pi ^{-1}$ from Eq. \eqref{appen-ILTYY-eq-7} and $\mathcal{H}_{\frac{1}{2}}(x) = \sqrt{\frac{2}{\pi}} ~e^{x}$ from Eq. \eqref{extreme-eq-8}. Plugging these forms in the expression of $\mathcal{G}_m^{\alpha}(z)$ in Eq. \eqref{extreme-res-eq-2} and performing the integration over $w$, we get
$$\mathcal{G}_m^{\alpha}(z) = \frac{1}{\pi\sqrt{z(1-z)}},~~~~~(\text{for } \alpha =0)$$
which matches with the \textit{arcsine law} in Eq. \eqref{extreme-eq-20-ne-2}. Curiously, the scaling function $\mathcal{G}_m^{\alpha}(z)$ is symmetric under the transformation $z \to 1-z$ only for $\alpha =0$. On the other hand, for non-zero $\alpha$, we see that $\mathcal{G}_m^{\alpha}(z)$ is not symmetric under this transformation. This is also exemplified in Figure \ref{timemax-fig}. To see this clearly, let us look at the form of $\mathcal{G}_m^{\alpha}(z)$ for $z \to 0$ and $z \to 1$. For general $\alpha$, we find (see Sec. V in SM \cite{Supplementary}) that the scaling function diverges as $\mathcal{G} _{m}^{\alpha} \left( z \to 0 \right) \sim z^{-\frac{1+\alpha}{2+\alpha}}$ and $\mathcal{G} _{m}^{\alpha} \left( z \to 1 \right) \sim (1-z)^{-\frac{1}{2}}$. Clearly, the divergences at two ends of $z$ are asymmetric for all $\alpha \neq 0$. Surprisingly, the divergence of the scaling function as $z \to 1$ is completely universal characterised by an $\alpha$-independent exponent $1/2$. The prefactor, however, may depend on the value of $\alpha$ as illustrated in Sec. V in SM \cite{Supplementary}. Heuristically, this $\alpha$-independence divergence can be understood as follows: From Eq. \eqref{extreme-eq-1}, we see that the joint distribution $\mathcal{P}(M, t_m|t)$ is proportional to the survival probability $S_M(t-t_m|M-\epsilon)$ which for $t_m \to t^-$ scales as $S_M(t-t_m|M-\epsilon) \sim  (t-t_m)^{-1/2}$ for all values of $\alpha$ [via Eq. \eqref{extreme-eq-10}]. Consequently, the marginal distribution $\mathcal{P}_m(t_m|t)$ also diverges as $(t-t_m)^{-1/2}$ for all values of $\alpha$.

\begin{figure}[t]
\includegraphics[scale=0.3]{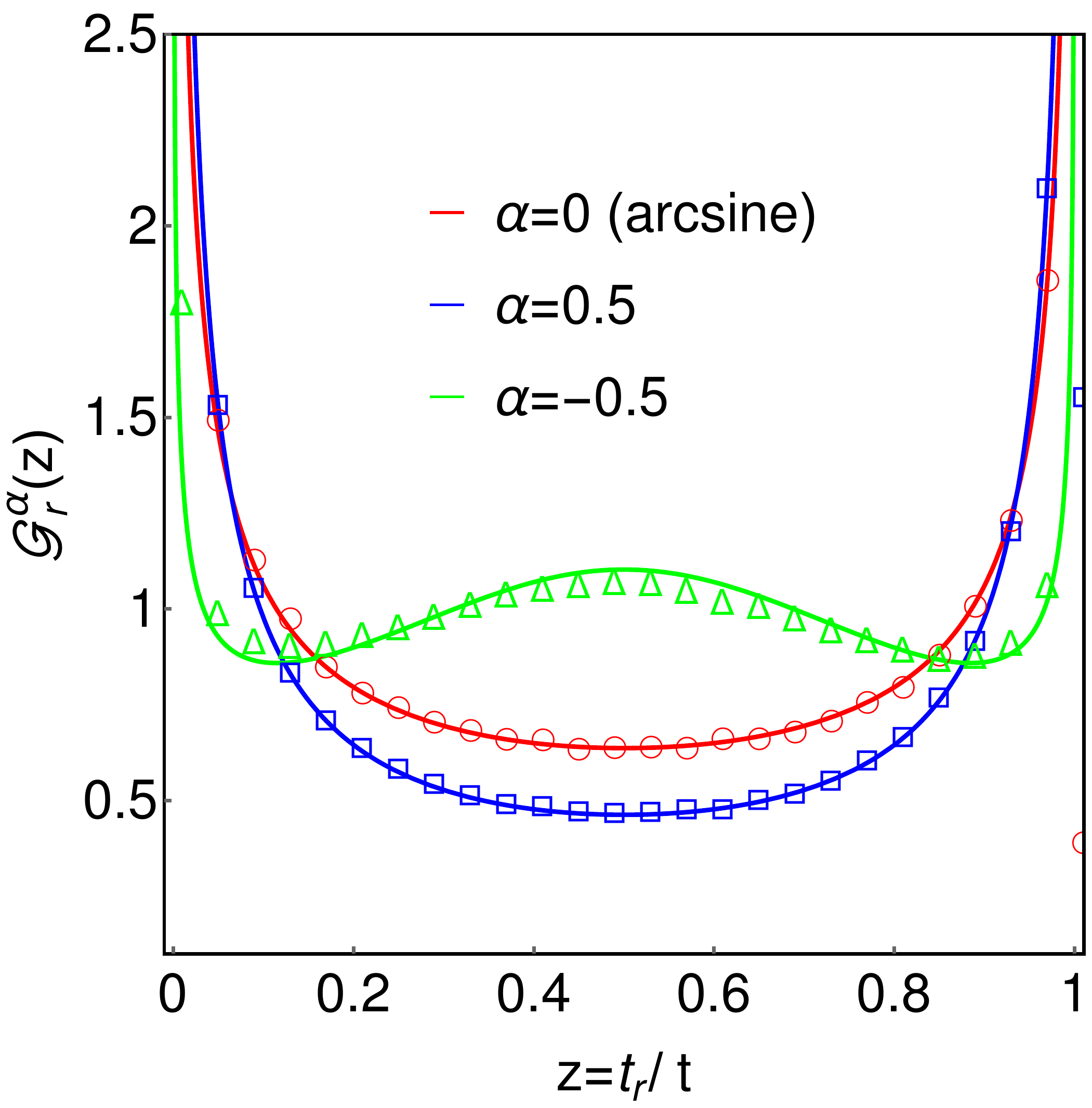}
\centering
\caption{Scaling function $\mathcal{G}_{r}^{\alpha}(z)$ in Eq. \eqref{resi-eq-9} for the occupation time distribution is plotted and compared with the numerical simulation for three different values of $\alpha$. The analytic expression in Eq. \eqref{resi-eq-9} is shown in solid line and the simulation data are shown by symbols. We have chosen $D_0=0.1$, $l=1$ and $t=5$. }
\label{occupation-fig}
\end{figure}
\section{Residence time distribution $\mathcal{P}_r(t_r|t)$}
\label{residence-time}
Residence time refers to the amount of time that the particle spends in the $x > 0$ region till duration $t$. Formally, it is defined as $t_r(t) = \int _{0}^{t} d\tau ~\Theta \left( x(\tau) \right)$, where $\Theta (x)$ denotes the Heaviside theta function. For standard Brownian motion, the distribution of $t_r$ is given in Eq. \eqref{eq-1}. This expression reveals that the distribution peaks (diverges) as $t_{r} \to  0^+$ and $t_{r} \to t^-$, whereas it exhibits minimum value at $t_r = t/2$. This implies rather a counter-intuitive property of the Brownian motion where once it crosses the origin on positive or negative side, it is reluctant to come back \bluew{\cite{Majumdar005}}. A natural question is: what happens to this property for general $\alpha$? In order to answer this question, we look at the residence time distribution for general $\alpha$ in this section.

Let us denote the distribution of $t_r$ by $\mathcal{P}_r(t_r, x_0|t)$ where $x_0$ is the initial position and $t$ is the total observation time. We later take $x_0 = 0$. Denoting the Laplace transformation of $\mathcal{P}_r(t_r, x_0|t)$ with respect to $t_r$ as $\mathcal{Q}(p,x_0|t)$, we have 
\begin{align}
\mathcal{Q}(p,x_0|t) &= \langle e^{-p t_r} \rangle , \label{resi-eq-1} \\
&= \int _{0}^{\infty} dt_r ~e^{-p t_r}~ \mathcal{P}_r(t_r, x_0|t).
\label{resi-eq-2}
\end{align}
The Laplace transform $\mathcal{Q}(p,x_0|t)$ satisfies the following backward master equations \bluew{\cite{Majumdar005}}:
\begin{align}
\partial _t \mathcal{Q}(p,x_0|t) = \left[D(x_0) \partial _{x_0}^2-p~ \Theta (x_0) \right]\mathcal{Q}(p,x_0|t),
\label{resi-eq-3}
\end{align}
where $D(x_0)$ is defined in Eq. \eqref{model-eq-2}. In order to solve this equation, we need to specify the appropriate initial and boundary conditions. For initial condition, we note that if $t \to 0$, then the residence time $t_r$ also tends to zero, i.e. $t_r \to 0$. Using this in Eq. \eqref{resi-eq-1}, we obtain
\begin{align}
\mathcal{Q}(p,x_0|t \to 0) = 1.
\label{resi-eq-4}
\end{align}
On the other hand, for any finite $t$, we have the following boundary conditions:
\begin{align}
& \mathcal{Q}(p,x_0 \to -\infty |t) = 1, \label{resi-eq-5} \\
& \mathcal{Q}(p,x_0 \to \infty |t) = e^{-pt}. \label{resi-eq-6}
\end{align}
Note that the first boundary condition follows from the fact that if $x_0 \to -\infty$, then the particle essentially stays in the $x<0$ region for all finite $t$. Consequently $t_r = 0$ which from Eq. \eqref{resi-eq-1} leads to $\mathcal{Q}(p,x_0 \to -\infty |t) = 1$. On the other hand, if $x_0 \to \infty$, then the particle stays in $x >0$ region for all finite $t$ and $t_r =t$. Plugging this in Eq. \eqref{resi-eq-1} results in the second boundary condition in Eq. \eqref{resi-eq-6}.

We now proceed to solve the backward equation \eqref{resi-eq-3} with these initial and boundary conditions. To this aim, we take another Laplace transformation of $\mathcal{Q}(p,x_0|t)$ with respect to $t$ and denote it by $\bar{\mathcal{Q}}(p,x_0|s)$. One can then appropriately transform the backward equation in terms of $\bar{\mathcal{Q}}(p,x_0|s)$ and solve it. To maintain continuity of the presentation, we have relegated these details to Sec. VI in SM \cite{Supplementary}. \greenw{The final solution for $x_0 = 0$ reads}
\begin{align}
\bar{\mathcal{Q}}(p|s) = \frac{1}{s}-\frac{p}{s(s+p)} \left[1+ \left( \frac{s}{s+p} \right) ^{\frac{1}{2+\alpha}} \right]^{-1},
\label{resi-eq-7}
\end{align}
where we have used the short hand notation $\bar{\mathcal{Q}}(p|s) = \bar{\mathcal{Q}}(p, x_0=0|s)$. One now has to perform the double inverse Laplace transformation of $\bar{\mathcal{Q}}(p|s)$ to get distribution in the time domain. \greenw{Fortunately, this inversion can be performed for all $\alpha>-1$ \bluew{\cite{Lamperti58, Carmi2010}}.} The distribution $\mathcal{P}_r(t_r|t)$ indeed has the scaling form in Eq. \eqref{resi-eq-8} and the scaling function $\mathcal{G} _{r}^{\alpha} \left( z \right)$ is given in Eq. \eqref{resi-eq-9}.

For $\alpha =0$, we recover the \textit{arcsine law} in Eq. \eqref{eq-1} for distribution $\mathcal{P}_r(t_r|t)$ in Eq. \eqref{resi-eq-8}. In Figure \ref{occupation-fig}, we have illustrated the scaling function $\mathcal{G} _{r}^{\alpha} \left( z \right)$ for three different values of $\alpha$ and compared them against the simulation. We find excellent agreement between them. Contrary to the arg-maximum $t_m(t)$, we see that $\mathcal{G} _{r}^{\alpha} \left( z \right)$ is symmetric about $z = \frac{1}{2}$ for all values of $\alpha$. In fact, from Eq. \eqref{resi-eq-9}, we see that it diverges as $z^{-\frac{1+\alpha}{2+\alpha}}$ and $(1-z)^{-\frac{1+\alpha}{2+\alpha}}$ as $z \to 0$ and $z \to 1$ respectively. Interestingly, in Figure \ref{occupation-fig}, we see that the scaling function exhibits local maxima at $z=1/2$ for $\alpha =-0.5$ which is in contrast to the other two values of $\alpha$ for which one finds minima at $z=1/2$. To understand this behaviour, we analyze $\mathcal{G} _{r}^{\alpha} \left( z \right)$ in the vicinity of $z = 1/2$. Expanding $\mathcal{G} _{r}^{\alpha} \left( z \right)$ in Eq. \eqref{resi-eq-9} for $z = \frac{1+ \bar{\epsilon}}{2}$ with $\bar{\epsilon} \to 0 $, we get
\begin{align}
& \mathcal{G} _{r}^{\alpha} \left( \frac{1+ \bar{\epsilon}}{2} \right) \simeq \frac{2}{\pi} \tan \left( \frac{\pi}{2(2+\alpha)}\right) +\frac{\mathbb{K}(\alpha)~ \bar{\epsilon} ^2}{2 \pi}, \label{resi-eq-10} 
\end{align}
 where the function $\mathbb{K}(\alpha)$ is defined as
\begin{align}
\mathbb{K}(\alpha) = \frac{\left[ 2+\alpha (4+\alpha) + (2+\alpha)^2 \cos \left( \frac{\pi}{2+\alpha}\right) \right]}{(2+\alpha)^2~\sin ^{-1}\left( \frac{\pi}{2+\alpha}\right)~\sec ^{-4} \left( \frac{\pi}{2(2+\alpha)}\right)}. 
\label{resi-eq-11}
 \end{align}
Now, $\mathcal{G} _{r}^{\alpha}(z) $ will exhibit local maxima or minima at $z = 1/2$ depending on whether $\mathbb{K}(\alpha)$ is positive or negative. Defining the critical value of $ \alpha $ as
\begin{align}
 \mathbb{K}(\alpha _c) = 0, \implies \alpha _c \simeq -0.3182,
\label{resi-eq-11}
\end{align}
we find that $\mathbb{K}(\alpha) >0$ for $\alpha \geq \alpha _c$ and $\mathbb{K}(\alpha) <0$ for $\alpha < \alpha _c$. Therefore, we expect a local maxima at $z=1/2$ for the scaling function $\mathcal{G} _{r}^{\alpha}(z) $ for $\alpha < \alpha _c$. Quite remarkably, this implies that for $\alpha \geq \alpha _c$, the particle, starting from the origin, typically stays entirely on the positive side or entirely on the negative side. The paths in which the particle spends equal amount of time on the positive and negative sides are relatively rare. This `stiff' property (or reluctance to cross the origin) has been long known for BM \bluew{\cite{Majumdar005}}. Here, we have shown that it gets extended for all $\alpha \geq \alpha _c$. On the other hand, for $\alpha < \alpha _c$, this `stiffness' is reduced which results in the local maximum of $\mathcal{G} _{r}^{\alpha}(z) $ at $z=1/2$. In fact, as $\alpha \to -1^+$, the scaling function simply becomes $\delta \left( z-1/2 \right)$. This implies that the particle typically spends equal amount of time on the positive and negative sides of the origin which is in sharp contrast to the standard Brownian motion.
\begin{figure}[t]
\includegraphics[scale=0.35]{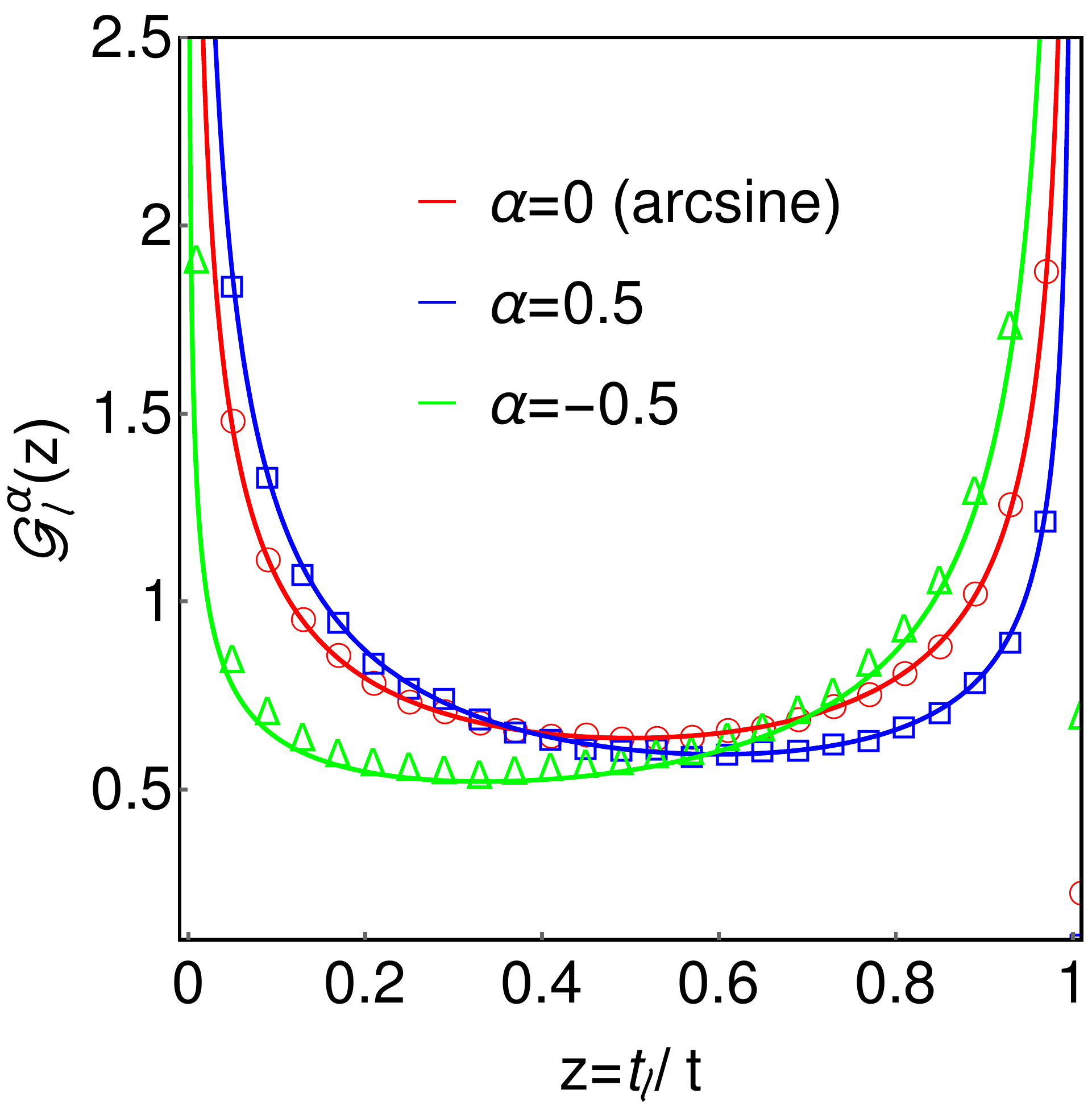}
\centering
\caption{Scaling function $\mathcal{G}_{\ell }^{\alpha}(z)$ in Eq. \eqref{last-passage-time-eq-9} for the last passage time distribution is plotted and compared with the numerical simulation for three different values of $\alpha$. The analytic expression in Eq. \eqref{last-passage-time-eq-9} is shown in solid line and the simulation data are shown by symbols. We have chosen $D_0=0.1$, $l=1$ and $t=5$. }
\label{last-passage-fig}
\end{figure}

\section{Last Passage time distribution $\mathcal{P}_{ \ell } \left( t_{\ell}|t \right)$}
\label{last-passage-time}
We now study the probability distribution $\mathcal{P}_{ \ell } \left( t_{\ell}|t \right)$ of time $t_{\ell}$ that the particle crosses orgin for the last time till duration $t$. As illustrated in Figure \ref{trajectory-pic-1}, we can analyse this problem by decomposing the trajectory into two parts. In the first part, particle reaches the orgin at time $t_{\ell}$ after starting its motion initially from the origin. The weight of this part is just the free probability distribution  $\mathbb{P}(0, t_{\ell}|0)$. In the second part, particle does not cross the origin in the remaining time interval $(t-t_{\ell})$ given that it was at the origin at time $t_l$. Then, the contribution of this part to $\mathcal{P}_{ \ell } \left( t_{\ell}|t \right)$ is the survival probability $S_0(t-t_{\ell}|0)$. Since the process is Markovian, these two contributions are statistically independent.

However, one encounters similar problem as encountered for the case of of extreme value statistics in sec. \ref{extreme-statistics}. Recall from this analysis of extreme value statistics that $S_0(t-t_{\ell}|0)$ is exactly equal to zero for all $t-t_{\ell}$ . In order to circumvent this problem, we instead compute the quantities $\mathbb{P}(\epsilon, t_{\ell}|0)$ and $S_0(t-t_{\ell}|\epsilon)$ and take $\epsilon \to 0^+$ limit at an appropriate stage of the calculation. Then, the distribution $\mathcal{P}_{ \ell } \left( t_{\ell}|t \right)$ can be written as
\begin{align}
\mathcal{P}_{ \ell } \left( t_{\ell}|t \right) = \frac{\mathbb{P}(\epsilon, t_{\ell}|0)~S_0(t-t_{\ell}|\epsilon)}{\mathcal{N}_L(\epsilon)},
\label{last-passage-time-eq-1}
\end{align}
where the function $\mathcal{N}_L(\epsilon)$ is just the normalisation factor. It is instructive to take the double Laplace transformation of this equation with respect to $t_{\ell}~(\to p)$ and $t~(\to s)$ to get
\begin{align}
\bar{\mathcal{P}}_{ \ell } \left( p|s \right) = \frac{\bar{\mathbb{P}}(\epsilon, s+p|0)~\bar{S}_0(s|\epsilon)}{\mathcal{N}_L(\epsilon)}. 
\label{last-passage-time-eq-2}
\end{align}
In this equation, we have used the notation $\bar{\mathcal{P}}_{ \ell } \left( p|s \right)$ and $\bar{\mathbb{P}}(\epsilon, s|0)$ to denote the Laplace transform of $\mathcal{P}_{ \ell } \left( t_{\ell}|t \right)$ and $\mathbb{P}(\epsilon, t|0)$ respectively. Interestingly, Eq. \eqref{last-passage-time-eq-2} implies that the problem of last-passage time has now been reduced to the problem of computing survival probability and distribution in an infinite line. The Laplace transforms $\bar{\mathbb{P}}(\epsilon, s|0)$ and $\bar{S}_0(s|\epsilon)$ can be explicitly obtained [see Sec. VII in SM \cite{Supplementary} for details] to be
\begin{align}
& \bar{\mathbb{P}}(\epsilon, s|0) \simeq \frac{\mathcal{A}_L (\epsilon)}{s^{\frac{1}{2+\alpha}}}, \label{last-passage-time-eq-3} \\
& \bar{S}_0(s|\epsilon) \simeq\frac{\mathcal{B}_L (\epsilon)}{s^{\frac{1+\alpha}{2+\alpha}}}, \label{last-passage-time-eq-4}
\end{align} 
where $\mathcal{A}_L (\epsilon)$ and $\mathcal{B}_L (\epsilon)$ are functions of $\epsilon$ whose explicit forms are given, respectively, in Eqs. (S105) and (S107) in SM \cite{Supplementary}. Next, we insert Eqs. \eqref{last-passage-time-eq-3} and \eqref{last-passage-time-eq-4} in Eq. \eqref{last-passage-time-eq-2} to write $\bar{\mathcal{P}}_{ \ell } \left( p|s \right)$ as
\begin{align}
\bar{\mathcal{P}}_{ \ell } \left( p|s \right) \simeq \frac{\mathcal{A}_L (\epsilon)~\mathcal{B}_L (\epsilon)}{\mathcal{N}_L(\epsilon)}~\frac{1}{s^{\frac{1+\alpha}{2+\alpha}} \left(s+p \right)^{\frac{1}{2+\alpha}}}.
\label{last-passage-time-eq-5}
\end{align}
We now have to specify the normalisation factor $\mathcal{N}_L(\epsilon)$. To evaluate this factor, we use the normalisation condition $\bar{\mathcal{P}}_{ \ell } \left( 0|s \right) = 1/s $ from which it is easy to show that $\mathcal{N}_L(\epsilon) =\mathcal{A}_L (\epsilon)~\mathcal{B}_L (\epsilon)$. This leads us to write $\bar{\mathcal{P}}_{ \ell } \left( p|s \right)$ as
\begin{align}
\bar{\mathcal{P}}_{ \ell } \left( p|s \right)= \frac{1}{s^{\frac{1+\alpha}{2+\alpha}} \left(s+p \right)^{\frac{1}{2+\alpha}}}.
\label{last-passage-time-eq-6}
\end{align}
\greenw{Finally, performing the double inverse Laplace transformation of this equation, we find that the distribution $\mathcal{P}_{ \ell } \left( t_{\ell}|t \right)$ of the last passage time $t_{\ell}$, for $\alpha >-1$, indeed possesses the scaling behaviour of Eq. \eqref{last-passage-time-eq-7} with the scaling function $\mathcal{G}_{\ell}^{\alpha}(z)$ defined in  Eq. \eqref{last-passage-time-eq-9} for general $\alpha$.} In Figure \ref{last-passage-fig}, we have plotted $\mathcal{G}_{\ell}^{\alpha}(z)$ for three values of $\alpha$ and compared against the numerical simulations. We observe excellent match for all $\alpha$.

One again, we see from Eq. \eqref{last-passage-time-eq-9} that $\mathcal{G}_{\ell}^{\alpha}(z)$ possesses $z \to 1-z$ symmetry only for $\alpha =0$. However, the symmetry is absent for non-zero values for  $\alpha$ as elucidated in Figure \ref{last-passage-fig}. Consequently, we get different divergences of the scaling function at the two ends, i.e.  $\mathcal{G} _{\ell }^{\alpha} \left( z \to 0 \right) \sim z^{-\frac{1+\alpha}{2+\alpha}}$ and $\mathcal{G} _{ \ell }^{\alpha} \left( z \to 1 \right) \sim (1-z)^{-\frac{1}{2+\alpha}}$. Physically, this asymmertic nature can be understood in the following way: For $ \alpha >0 $, the particle typically stays away from the origin due to the large values of the diffusion coefficient around the origin. This gives rise to the smaller values of $t_{ \ell }$. As a result, the distribution $\mathcal{P}_{ \ell } \left( t_{\ell}|t \right)$ is sharply peaked at the smaller values of $t_{\ell}$. On the other hand, for $\alpha <0$, the particle typically stays near the origin which enhances its chances to cross the origin. This essentially gives rise to the large values of $t_{\ell}$ and peaking of $\mathcal{P}_{ \ell } \left( t_{\ell}|t \right)$ at these values.

\section{CONCLUSION}
\label{conclusion}
To conclude, we have studied a model of anomalous diffusion in which a single particle moves in an one dimensional heterogeneous medium with spatially varying diffusion coefficient of the form $D(x) \sim |x|^{-\alpha}$ with $\alpha >-1$. Depending on the exponent $\alpha$, this model displays super-diffusive $(-1<\alpha <0)$, diffusive $(\alpha =0)$ or sub-diffusive $(0<\alpha <\infty)$ scaling of the mean-squared displacement (MSD). Curiously, this simple Markov process also exhibits weak ergodicity breaking in the sense that the time-averaged and ensemble averaged MSDs are not equal even at large times \cite{Cherstvy2013, Cherstvy2014, Leibovich2019,Cherstvy2015, Wang2019, Fa2005, new2}.

In this paper, we extensively investigated the statistical properties of the maximum displacement $M(t)$ and time $t_m(t)$ taken to reach this maximum till duration $t$. Exploiting the path decomposition technique for Markov processes \bluew{\cite{tmax-2}}, we derived, for all $\alpha$, the joint probability distribution of $M(t)$ and $t_m(t)$. Marginalising this joint distribution for $M(t)$ shows that the distribution $P_m(M|t)$ possesses scaling behaviour in $M/t^{\frac{1}{2+\alpha}}$ with the corresponding scaling function $\mathcal{F}_{\alpha}(z)$ rigorously derived in Eq. \eqref{extreme-eq-18}. \greenw{Contrary to the standard Brownian motion (BM), we obtain that $\mathcal{F}_{\alpha}(z)$, for non-zero $\alpha$, has a non-monotonic dependence on $z$ which is a consequence of the heterogeneity in the environment. Behaviour of $\mathcal{F}_{\alpha}(z)$ for $\alpha <0$ and $\alpha >0$ is also rather different. For $\alpha <0$, the scaling function rises initially for small $z$, attains a maximum value and then decays for large $z$. On the other hand, for $\alpha >0$, it initially decreases with $z$ then rises at some intermediate $z$ until it attains a local maximum. After that, it again decreases for large $z$. This behaviour has been illustrated in Figure \ref{scaling-M-fig}.} 

Our analysis on extreme-value statistics also provides an exact expression of the marginal distribution $\mathcal{P}_m(t_m|t)$ of the arg-maximum $t_m(t)$ in Eq. \eqref{extreme-res-eq-1}. \greenw{In contrast to the BM, we find that the distribution $\mathcal{P}_m(t_m|t)$, for $\alpha \neq 0$, is not symmetric about $t_m = t/2$. This difference is also exemplified by asymmetric peaks (divergences) of $\mathcal{P}_m(t_m|t)$ as $t_m \to 0^+$ and $t_m \to t^-$, namely, $\mathcal{P}_m \left( t_m \to 0|t \right) \sim t_m^{-\frac{1+\alpha}{2+\alpha}}$ and  $\mathcal{P}_m \left( t_m \to t|t \right) \sim (t-t_m)^{-\frac{1}{2}}$. Recall that for BM, $\mathcal{P}_m(t_m|t)$ is symmetric about $t_m = t/2$ and diverges identically as $\mathcal{P}_m \left( t_m \to 0|t \right) \sim t_m^{-\frac{1}{2}}$ and  $\mathcal{P}_m \left( t_m \to t|t \right) \sim (t-t_m)^{-\frac{1}{2}}$.}

The second part of our paper dealt with the analysis of the residence time $t_r(t)$ for which we computed the probability distribution $\mathcal{P}_r \left( t_r|t\right)$ exactly for all values of $\alpha$. Quite remarkably, we find the existence of a critical $\alpha$ (which we denote by $\alpha _c = -0.3182$) such that $\mathcal{P}_r \left( t_r|t\right)$ has minima at $t_r = t/2$ for $\alpha \geq \alpha _c$ whereas it exhibits local maximum at $t_r = t/2$ for $\alpha < \alpha _c$. We also provided a simple physical reasoning of this behaviour based on the likelihood of the particle to stay on one side of the origin. \greenw{Appearance of local maxima at $t_r=t/2$ is in sharp contrast to the standard BM.} Finally, we calculated the distribution  $\mathcal{P}_{\ell} \left( t_{\ell}|t\right)$ of the last-passage time $t _{\ell}(t)$ and showed that it is also asymmetric about $t _{\ell}(t) = t/2$ for non-zero $\alpha$. This is further illustrated by the difference in behaviour of $\mathcal{P}_{\ell} \left( t_{\ell}|t\right)$ as $t _{\ell}  \to 0^+$ and $t _{\ell}  \to t^-$, viz. $\mathcal{P}_{\ell} \left( t_{\ell} \to 0|t\right)\sim t_{\ell}^{-\frac{1+\alpha}{2+\alpha}} $ and $\mathcal{P}_{\ell} \left( t_{\ell} \to t|t\right)\sim(t- t_{\ell})^{-\frac{1}{2+\alpha}} $. \greenw{We emphasize that while the distributions of $t_m(t),~t_r(t)$ and $t _{\ell}(t)$ are all identical to Eq. \eqref{eq-1} for $\alpha =0$ , they turn out to be significantly different for $\alpha \neq 0$.} In fact, for $\alpha =0$ (BM), the equivalence between $t_m(t)$ and  $t _{\ell}(t)$ can be established based on the reflection property, inversion symmetry and time reversal symmetry \bluew{\cite{Feller}}. However, these symmetries are not present for $\alpha \neq 0$ which results in inequivalence between $t_m(t)$ and  $t _{\ell}(t)$.

Here, we have showcased a simple example of heterogeneous diffusion model driven by white Gaussian noise for which we could derive many results on extremal statistics and path functionals exactly. Unravelling these results for other complex heterogeneous models remains a promising future direction. Recently heterogeneous diffusion processes driven by coloured noise have garnered significant interest due to their potential application in biological systems \cite{MutothyaXu2021,Mutothya2021,Xu2020}. It would be interesting to see how our results get modified in these scenarios. \greenw{Another interesting direction is to explore the ramifications of the combined effect of HDP and other models like fractional Brownian motion \cite{tmax-FBM-3,tmax-FBM-1,new1, Rnew2} and scaled Brownian motion \cite{Rnew1} on the extreme-value statistics and arcsine laws.} 

\greenw{Finally, we remark that our work may be verified in experiments involving diffusion of tracer proteins in the cytoplasmic part of the cell where substantial heterogeneity arises due to the non-uniform distribution of various crowding obstacles like ribosomes, nuclei acids, cytoskeleton \cite{Kuhn2011}. Space-dependent diffusion coefficient is also observed in experiments involving particles trapped between two nearly-parallel plates \cite{Lancon2001}. It would be interesting to compare our analytical results with these experiments.}

\section{Acknowledgement}
I am indebted to my supervisor Dr. Anupam Kundu for collaboration on other projects upon which the current work is based. I acknowledge support of the Department of Atomic Energy, Government of India, under project no.12-R\&D-TFR-5.10-1100.

\appendix

\section{Derivation of $\bar{P}(M, p|s)$ in Eq. \eqref{extreme-eq-2}}
\label{appen-exp}
Here, we derive the expression of the Laplace transform $\bar{P}(M, p|s)$ of joint distribution in Eq. \eqref{extreme-eq-12}. From 
Eq. \eqref{extreme-eq-2}, we see that this reduces to the problem of computing $\bar{S}_M(s|M-\epsilon)$ and $\bar{F}_{M-\epsilon}\left(s+p|0 \right)$. Using Eq. \eqref{extreme-eq-7}, we get
\begin{align}
& \bar{S}_M(s|M-\epsilon) \simeq \frac{\epsilon~ \partial _M  \left[\mathcal{H}_{\frac{1}{2+\alpha}} \left( (a_s M)^{\frac{2+\alpha}{2}} \right) \right]}{s ~\mathcal{H}_{\frac{1}{2+\alpha}} \left( (a_s M)^{\frac{2+\alpha}{2}} \right) }, \label{extreme-eq-10}\\
& \bar{F}_{M-\epsilon}\left(s+p|0 \right) = 1-(s+p)~\bar{S}_{M-\epsilon}(s+p|0),  \nonumber \\
& ~~~~~~~~~~~~~~~~~~~~\simeq \frac{\mathcal{H}_{\frac{1}{2+\alpha}} \left(0 \right)}{\mathcal{H}_{\frac{1}{2+\alpha}} \left( (a_{s+p} M)^{\frac{2+\alpha}{2}} \right)}.
\label{extreme-eq-11}
\end{align}
Finally, inserting Eqs. \eqref{extreme-eq-10} and \eqref{extreme-eq-11} in Eq. \eqref{extreme-eq-2} results
\begin{align}
\bar{P}(M, p|s)&= \frac{ \epsilon ~ \mathcal{H}_{\frac{1}{2+\alpha}} \left(0 \right) }{\mathcal{N} (\epsilon)s~\mathcal{H}_{\frac{1}{2+\alpha}} \left( (a_{s+p} M)^{\frac{2+\alpha}{2}} \right)}
\nonumber  \\
& ~~~~~~\times \frac{\partial _M \left[ \mathcal{H}_{\frac{1}{2+\alpha}} \left( (a_s M)^{\frac{2+\alpha}{2}} \right) \right]}{ \mathcal{H}_{\frac{1}{2+\alpha}} \left( (a_s M)^{\frac{2+\alpha}{2}} \right) } . 
\end{align}
This result has been quoted in Eq. \eqref{extreme-eq-12}.
\begin{widetext}
\section{Important Formulae}
\label{appen-ILT-J}
We present here a list of the expressions and the notations that we have used in our paper:
\begin{align}
& \mathcal{H}_{\beta} (w) = w ^{\beta} \left[ I _{\beta}(w)+I_{-\beta} (w)\right],\\
& \mathbb{H}_{\beta}(w)= \frac{1}{2 \pi \beta i} \left[\frac{1}{\mathcal{H}_{\beta} \left( -i w \right)}-\frac{1}{\mathcal{H}_{\beta} \left( i w \right)} \right].
\label{appen-ILT-J-Eq-6} \\
& \mathbb{X}_{\alpha} \left( w\right) =\frac{e^{\frac{i \pi }{2+\alpha}}}{2 \pi i} \left[\frac{\mathcal{H}_{\frac{1+\alpha}{2+\alpha}} \left( \frac{i}{{w}}\right)}{\mathcal{H}_{\frac{1}{2+\alpha}} \left( \frac{i}{{w}}\right)} -\frac{e^{-\frac{2i \pi }{2+\alpha}}~\mathcal{H}_{\frac{1+\alpha}{2+\alpha}} \left( -\frac{i}{{w}}\right)}{\mathcal{H}_{\frac{1}{2+\alpha}} \left(- \frac{i}{{w}}\right)} \right].
\label{appen-ILTYY-eq-7}
\end{align}
\end{widetext}


\end{document}


	
	
	\date{\today}
	\newcommand{\titlename}{Supplementary material to ``Extreme value statistics and arcsine laws for heterogeneous diffusion processes"}

	\title{\titlename}
	
\author{Prashant Singh}
\email{prashant.singh@icts.res.in}
\affiliation{International Centre for Theoretical Sciences, Tata Institute of Fundamental Research, Bengaluru 560089, India}
\maketitle
	
	
This supplementary material (SM) provides a detailed derivation of some results which were used in the main text. For self-containedness, let us recall our model of heterogeneous diffusion processes (HDP) where the position of the particle evolves as
\begin{align}
\frac{dx}{dt} = \sqrt{2 D(x)} ~\eta(t),
\label{new-appen-model-eq-1}
\end{align}
where $\eta(t)$ is the Gaussian white noise with zero mean and correlation $\langle \eta(t) \eta(t') \rangle = \delta(t-t')$. We focus on the power-law form of the diffusion coefficient:
\begin{align}
D(x) =  \frac{D_0~l^{\alpha}}{|x|^{\alpha}},
\label{new-appen-extreme-eq-3}
\end{align}
where $D_0$ is a positive constant that sets the strength of the noise and $l$ is the length scale over which $D(x)$ changes. The exponent $\alpha$ quantifies the strength of the gradient of $D(x)$. Throughout this SM, we consider $\alpha >-1$ and interpret Eq. \eqref{new-appen-model-eq-1} in Ito-setup.
	
\section{Derivation of the survival probability $S_M(t|x_0)$}
\label{appen-surv}
In this section, we derive the expression of the survival probability $S_M(t|x_0)$ which is significant in computing the extremal statistics of the model. To this aim, we begin with the backward Fokker Planck equation for $S_M(t|x_0)$ in Ito-set up \cite{newRedner}
\begin{align}
 \partial _t S_M(t|x_0) & = D(x_0) \partial _{x_0 } ^2 S_M(t|x_0). 
\end{align} 
Taking Laplace transformation of this equation with respect to $t ~(\to s)$ yields
\begin{align}
s \bar{S}_M(s|x_0)-1 = D(x_0) \partial _{x_0 }^2 \bar{S}_M(s|x_0).
\label{extreme-eq-6}
\end{align}
Next, we perform the transformation
\begin{align}
\bar{S}_M(s|x_0)=\frac{1}{s}+U(x_0),
\label{appen-surv-eq-1}
\end{align}
in Eq. \eqref{extreme-eq-6} which results in a homogeneous differential equation of the form
\begin{align}
s U(x_0) = D(x_0) \partial _{x_0 }^2 U(x_0),
\label{appen-surv-eq-2}
\end{align}
with $D(x_0)$ defined in Eq. \eqref{new-appen-extreme-eq-3}. Recall that $M \geq 0$ and $x_0 \leq M$. We proceed to solve Eq. \eqref{appen-surv-eq-2} separately for $x_0 >0$ and $x_0 <0$ regions. For $x_0 >0$, we perform the transformation $y = \frac{x_0 ^{2+\alpha}}{(2+\alpha)^2 l^{\alpha} D_0}$ and rewrite Eq. \eqref{appen-surv-eq-2} as
\begin{align}
y \frac{\partial ^2 U}{\partial y^2} + \left(\frac{1+\alpha}{2+\alpha} \right) \frac{\partial  U}{\partial y} = s U.
\label{appen-surv-eq-3}
\end{align}
The solution of this equation is given in terms of the modified Bessel functions as
\begin{align}
U(y) = \mathbb{C}_1 y^{\frac{1}{2(2+\alpha)}} K_{\frac{1}{2+\alpha}} \left( 2 \sqrt{s y}\right)+\mathbb{C}_2 y^{\frac{1}{2(2+\alpha)}} I_{\frac{1}{2+\alpha}} \left( 2 \sqrt{s y}\right).
\label{appen-surv-eq-4}
\end{align}
Here $\mathbb{C}_1$ and $\mathbb{C}_2$ are constants independent of $y$ but may depend on $s$. Writing the solution in terms of $x_0$ and using Eq. \eqref{appen-surv-eq-1} to write $\bar{S}_M(s|x_0)$ in terms of $U(x_0)$, we get
\begin{align}
\bar{S}_M(s|x_0) &= \frac{1}{s} +\mathbb{C}_1 \sqrt{x_0}K_{\frac{1}{2+\alpha}} \left(  (a_s x_0)^{\frac{2+\alpha}{2}}\right)+\mathbb{C}_2 \sqrt{x_0}I_{\frac{1}{2+\alpha}} \left(  (a_s x_0)^{\frac{2+\alpha}{2}}\right).
\label{appen-surv-eq-5}
\end{align} 
where the function $a_s$ is given by
\begin{align}
a_s  = \left(\frac{s}{\mathcal{D}_{\alpha}} \right)^{\frac{1}{2+\alpha}},~~~\text{with }\mathcal{D}_{\alpha} = \frac{D_0 l^{\alpha} (2+\alpha)^2}{4}.\label{new-appen-extreme-eq-9}
\end{align}
Recall that the solution in Eq. \eqref{appen-surv-eq-5} is true only for $x_0 >0$. For $x_0 <0$, we proceed similarly to get
\begin{align}
\bar{S}_M(s|x_0) = &\frac{1}{s} +\mathbb{C}_3 \sqrt{|x_0|}K_{\frac{1}{2+\alpha}} \left(  (a_s |x_0|)^{\frac{2+\alpha}{2}}\right) +\mathbb{C}_4 \sqrt{|x_0|}I_{\frac{1}{2+\alpha}} \left(  (a_s |x_0|)^{\frac{2+\alpha}{2}}\right).
\label{appen-surv-eq-6}
\end{align}
The task now is to evaluate the constants $\mathbb{C}_1,~\mathbb{C}_2,~\mathbb{C}_3$ and $\mathbb{C}_4$ which are, in principle, functions of $s$. To compute them, we first note that the survival probability $S_M(t|x_0)$ and its derivative $\partial _{x_0} S_M(t|x_0)$ are continuous across $x_0 = 0$ which implies that $\bar{S}_M(s|x_0)$ and $\partial _{x_0} \bar{S}_M(s|x_0)$ are also continuous. This implies
\begin{align}
& \bar{S}_M(s|x_0 \to 0^+) = \bar{S}_M(s|x_0 \to 0^-), \label{appen-surv-eq-7} \\
& \left[\frac{\partial \bar{S}_M(s|x_0 )}{\partial x_0}\right] _{x_0 \to 0^+} = \left[\frac{\partial \bar{S}_M(s|x_0 )}{\partial x_0}\right] _{x_0 \to 0^-}. \label{appen-surv-eq-8}
\end{align}
\greenw{Note that continuity of the derivative in the second equation is well-defined only for $\alpha >-1$ as can be verified explicitly by plugging the solutions of $\bar{S}_M(s|x_0)$ from Eqs. \eqref{appen-surv-eq-5} and \eqref{appen-surv-eq-6}.} Next, we recall the boundary conditions of $S_M(t|x_0)$ in Eqs. (20) and (21) in the main text. Translating these conditions in terms of $\bar{S}_M(s|x_0 )$ yields
\begin{align}
& \bar{S}_M(s|x_0 \to M) = 0,  \label{appen-surv-eq-9} \\
& \bar{S}_M(s|x_0 \to -\infty) = \frac{1}{s} \label{appen-surv-eq-10}.
\end{align}
Using the four conditions [Eqs. \eqref{appen-surv-eq-7}-\eqref{appen-surv-eq-10}], we obtain the constants $\mathbb{C}_1,~\mathbb{C}_2,~\mathbb{C}_3$ and $\mathbb{C}_4$. This, then, completely specifies the Laplace transform $\bar{S}_M(s|x_0)$ in Eqs. \eqref{appen-surv-eq-5} and \eqref{appen-surv-eq-6}. However, since we are eventually interested in computing the joint distribution of the maximum $M(t)$ and arg-max $t_m(t)$ [Eq. (17) in the main text], we provide the solution only for $x_0 \geq 0$. As evident from Eq. \eqref{appen-surv-eq-5}, we then need only the expressions of $\mathbb{C}_1$ and $\mathbb{C}_2$ which read
\begin{align}
& \mathbb{C}_1 = \frac{\mathbb{C}_2}{\Gamma \left(\frac{1}{2+\alpha} \right) \Gamma \left(\frac{1+\alpha}{2+\alpha} \right)}, \label{appen-surv-eq-11} \\
& \mathbb{C}_2 = -\frac{1}{s f _{\alpha}(M)},~~~~~\text{with },\label{appen-surv-eq-12} \\
&f _{\alpha}(M) = \frac{1}{2\sqrt{a_s}} \mathcal{H}_{\frac{1}{2+\alpha}} \left( (a_s M)^{\frac{2+\alpha}{2}} \right),
\end{align}
where $\mathcal{H}_{\beta}(x_0)$ in the last equation is defined as
\begin{align}
\mathcal{H}_{\beta} (x_0) = x_0 ^{\beta} \left[ I _{\beta}(x_0)+I_{-\beta} (x_0)\right]. \label{new-appen-extreme-eq-8} 
\end{align}
Finally, inserting these expressions of $ \mathbb{C}_1$ and $ \mathbb{C}_2$ in Eq. \eqref{appen-surv-eq-5}, we find 
\begin{align}
\bar{S}_M(s|x_0) = \frac{1}{s} \left[1- \frac{\mathcal{H}_{\frac{1}{2+\alpha}} \left( (a_s x_0)^{\frac{2+\alpha}{2}} \right)}{\mathcal{H}_{\frac{1}{2+\alpha}} \left( (a_s M)^{\frac{2+\alpha}{2}} \right)} \right],
\label{appen-surv-eq-13} 
\end{align}
which has been written in Eq. (22) in the main text.

\section{Marginal distribution $P_m(M|t)$}
\label{new-appen-ILT-J}
\begin{figure}[t]
\includegraphics[scale=0.4]{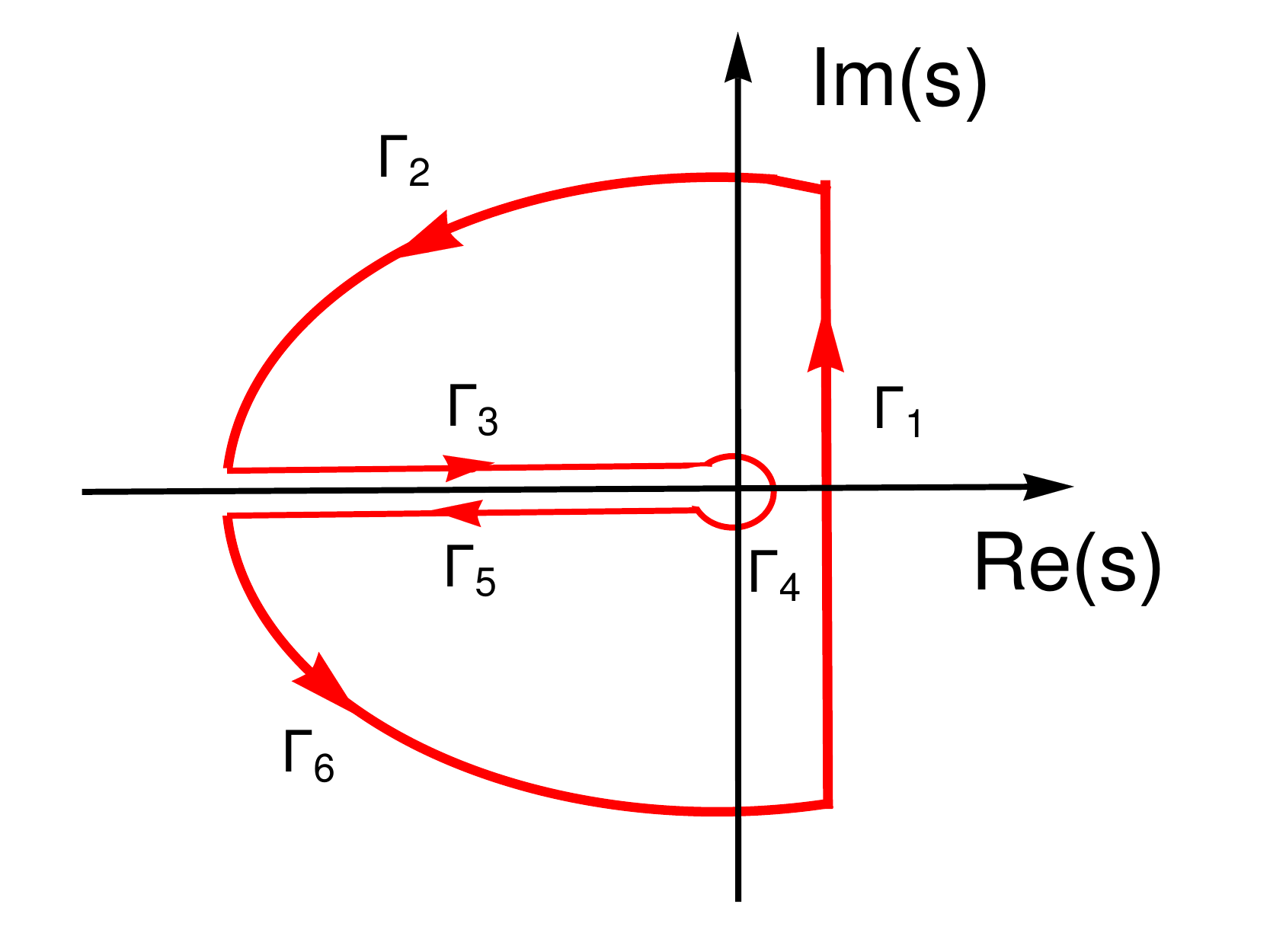}
\centering
\caption{Contour for the Bromwich integration of $\bar{J}(M,s)$ in Eq. \eqref{appen-ILT-J-Eq-1}}
\label{contour-fig}
\end{figure}

Let us now derive the exact form of the Marginal distribution $P_m(M|t)$ for the maximum $M$. In Eq. (28) of the main text, we obtained the Laplace transform $\bar{P}_m(M|s)$ of the marginal distribution $M$ to be
\begin{align}
\bar{P}_m(M|s)&  =-\frac{d \bar{J}(M,s)}{dM},~~~~~~~~~~~~\text{with} \label{new-appen-extreme-eq-15} \\
\bar{J}(M,s)&= \frac{\mathcal{H}_{\frac{1}{2+\alpha}} \left(0 \right)}{s ~\mathcal{H}_{\frac{1}{2+\alpha}} \left( (a_s M)^{\frac{2+\alpha}{2}} \right)}, \label{new-appen-extreme-eq-16} 
\end{align}
Here, we perform the inverse Laplace transformation of $\bar{P}_m(M|s)$ to obtain the distribution $P_m(M|t)$ as quoted in Eq. (8) of the main text. From Eq. \eqref{new-appen-extreme-eq-15}, we observe that $\bar{P}_m(M|s)$ is simply the derivative of $\bar{J}(M,s)$. In the time domain, this will correspond to
\begin{align}
P_m(M|t) = - \frac{dJ(M,t)}{dM},
\label{appen-ILT-J-Eq-0}
\end{align}
where $J(M,t)$ is the inverse Lapalace transform of $\bar{J}(M,s)$. We therefore proceed to compute the inverse Laplace transformation of  $\bar{J}(M,s)$ in Eq. \eqref{new-appen-extreme-eq-16}. Formally, $J(M,t)$ can be written as
\begin{align}
J(M,t) &= \frac{1}{2 \pi i} \int _{-i \infty}^{i \infty} ds~ e^{st}~\bar{J}(M,s), \\
& = \frac{\mathcal{H}_{\frac{1}{2+\alpha}} \left(0 \right)}{2 \pi i} \int _{-i \infty}^{i \infty} ds \frac{e^{st}}{s ~\mathcal{H}_{\frac{1}{2+\alpha}} \left( \sqrt{\frac{s M^{2+\alpha}}{\mathcal{D}_{\alpha}}}  \right)}.
\label{appen-ILT-J-Eq-1}
\end{align}
Notice that $s=0$ is a branch point. To perform this Bromwich integration, we consider the contour of form shown in Figure \ref{contour-fig}. Since the integrand is analytic inside this contour, the Cauchy theorem gives
\begin{align}
\int _{\Gamma _1}+\int _{\Gamma _2}+\int _{\Gamma _3}+\int _{\Gamma _4}+\int _{\Gamma _5}+\int _{\Gamma _6} = 0.\label{appen-ILT-J-Eq-2-new}
\end{align}
$\int _{\Gamma _1} = J(M,t)$ is the integral that we need. Let us now perform these integrals along different paths separately. Recall that the real part of $s$ along $\Gamma _2$ and $\Gamma _6$ is negative and in the limit $|s| \to \infty$, the contribution becomes exactly zero.

To evaluate integral along $\Gamma _3$, we substitute $s = R e^{i \pi}$, where $R$ varies from $\infty$ to $0$. Under this transformation, the integral $\int _{\Gamma _3}$ becomes
\begin{align}
\int _{\Gamma _3} = \frac{\mathcal{H}_{\frac{1}{2+\alpha}} \left(0 \right)}{2 \pi i} \int _{\infty}^{0} dR \frac{e^{-Rt}}{R ~\mathcal{H}_{\frac{1}{2+\alpha}} \left( i\sqrt{\frac{R M^{2+\alpha}}{\mathcal{D}_{\alpha}}}  \right)}.
\label{appen-ILT-J-Eq-2}
\end{align}
To simplify further, we substitute $w = \frac{R M^{2+\alpha}}{\mathcal{D}_{\alpha}}$ in this equation and get
\begin{align}
\int _{\Gamma _3} = -\frac{\mathcal{H}_{\frac{1}{2+\alpha}} \left(0 \right)}{2 \pi i} \int _{0}^{\infty} dw \frac{e^{-\frac{\mathcal{D}_{\alpha} t}{M^{2+\alpha}}w}}{w ~\mathcal{H}_{\frac{1}{2+\alpha}} \left( i \sqrt{w} \right)}.
\label{appen-ILT-J-Eq-3}
\end{align}
Next, we evaluate the integral along $\Gamma _5$. For this, we substitute $s = e^{-i \pi} R$ in the integrand and perform similar simplications as done for the integral $\int _{\Gamma _3}$. We then obtain
\begin{align}
\int _{\Gamma _5} = \frac{\mathcal{H}_{\frac{1}{2+\alpha}} \left(0 \right)}{2 \pi i} \int _{0}^{\infty} dw \frac{e^{-\frac{\mathcal{D}_{\alpha} t}{M^{2+\alpha}}w}}{w ~\mathcal{H}_{\frac{1}{2+\alpha}} \left( -i \sqrt{w} \right)}.
\label{appen-ILT-J-Eq-4}
\end{align}
Finally, we perform the integration along $\Gamma _4$ for which we replace $s=\delta e^{i \theta}$ and take $\delta \to 0^+$ limit. It is easy to show that the resultant integration is $\int _{\Gamma _4} = -1$.

Putting all the intergations together in Eq. \eqref{appen-ILT-J-Eq-2-new} and noting that $\int _{\Gamma _1} = J(M,t)$, we get
\begin{align}
J(M,t) &= -\int _{\Gamma _3}-\int _{\Gamma _4}-\int _{\Gamma _5}, \\
& = 1-\frac{\mathcal{H}_{\frac{1}{2+\alpha}} \left(0 \right)}{(2+\alpha)} \int _{0}^{\infty} \frac{dw}{w} e^{-\frac{\mathcal{D}_{\alpha} t}{M^{2+\alpha}}w}~\mathbb{H}_{\frac{1}{2+\alpha}}(\sqrt{w}),
\label{appen-ILT-J-Eq-5}
\end{align}
where the function $\mathbb{H}_{\beta}(w)$ is given by
\begin{align}
\mathbb{H}_{\beta}(w)= \frac{1}{2 \pi \beta i} \left[\frac{1}{\mathcal{H}_{\beta} \left( -i w \right)}-\frac{1}{\mathcal{H}_{\beta} \left( i w \right)} \right].
\label{appen-ILT-J-Eq-6}
\end{align}
Coming back to our goal of computing the distribution $P_m(M|t)$, we plug $J(M,t)$ from Eq. \eqref{appen-ILT-J-Eq-5} in Eq. \eqref{appen-ILT-J-Eq-0} to get
\begin{align}
P_m(M|t) = \frac{1}{\left( \mathcal{D}_{\alpha} t\right)^{\frac{1}{2+\alpha}}} \mathcal{F}_{\alpha} \left( \frac{M}{\left( \mathcal{D}_{\alpha} t\right)^{\frac{1}{2+\alpha}}} \right),
\label{new-appen-ILT-J-Eq-8} 
\end{align}
with $\mathcal{D}_{\alpha}$ given in Eq. \eqref{new-appen-extreme-eq-9} and the function $\mathcal{F}_{\alpha}(z)$ defined asw
\begin{align}
 \mathcal{F}_{\alpha}(z) = \frac{\mathcal{H}_{\frac{1}{2+\alpha}} \left(0 \right)}{z^{3+\alpha}} \int _{0}^{\infty} dw ~e^{-\frac{w}{z^{2+\alpha}}} ~\mathbb{H}_{\frac{1}{2+\alpha}}(\sqrt{w}), \label{new-appen-extreme-eq-18} 
\end{align}
where $\mathcal{H}_{\beta}(x_0)$ is in Eq. \eqref{new-appen-extreme-eq-8}. 

\section{Asymptotic behaviour of the scaling function $\mathcal{F}_{\alpha}(z)$}
\label{appen-asy-PM}
We now proceed to analyse the asymptotic behaviour of the scaling function  $\mathcal{F}_{\alpha}(z)$ in Eq. \eqref{new-appen-extreme-eq-18}. For simplicity, we look at the small and large $z$ behaviours separately below.
\subsection{Small $z$ behaviour of $\mathcal{F}_{\alpha}(z)$}
To obtain $\mathcal{F}_{\alpha} \left(z \to 0 \right)$, we consider the form of $\bar{P}(M|s)$ in terms of $\bar{J}(M,s)$ as shown in Eq. \eqref{new-appen-extreme-eq-15} and look at the small $M$ behaviour of $\bar{J}(M,s)$. As evident from Eq. \eqref{new-appen-extreme-eq-16}, one then needs to specify the  small-$x$ form of $\mathcal{H}_{\beta}(x)$. For $x \to 0$, the modified Bessel function is
\begin{align}
I _{\beta}(x) \simeq \frac{x^{\beta}}{2^{\beta} \Gamma(1+\beta)} + \frac{x^{2+\beta}}{2^{2+\beta}\Gamma(2+\beta)},
\label{appen-asy-PM-eq-1}
\end{align}
which we use in Eq. \eqref{new-appen-extreme-eq-8} to get
\begin{align}
\mathcal{H}_{\beta}(x) \simeq \mathcal{H}_{\beta} (0) +\frac{x^{2 \beta}}{2^{\beta} \Gamma(1+\beta)},~~~\text{as } x \to 0.
\label{appen-asy-PM-eq-2}
\end{align}
Substituting this form of $\mathcal{H}_{\beta}(x)$ in $\bar{J}(M,s)$ in Eq. \eqref{new-appen-extreme-eq-16}, we get 
\begin{align}
\bar{J}(M,s) \simeq \frac{C_{\alpha} \mathcal{D}_{\alpha} ^{\frac{1}{2+\alpha}}}{s \left( C_{\alpha} \mathcal{D}_{\alpha} ^{\frac{1}{2+\alpha}} +M s^{\frac{1}{2+\alpha}}\right)},~~~\text{as }M \to 0,
\label{appen-asy-PM-eq-3}
\end{align}
with $C_{\alpha} = \frac{2^{2/2+\alpha}}{(2+\alpha)} \frac{\Gamma \left( \frac{1}{2+\alpha}\right)}{\Gamma \left( \frac{1+\alpha}{2+\alpha}\right)}$. Next, we insert Eq. \eqref{appen-asy-PM-eq-3} in Eq. \eqref{new-appen-extreme-eq-15} to yield the small-$M$ behaviour of $\bar{P}_m(M|s)$ for $\alpha \neq 0$ as
\begin{align}
\bar{P}_m(M|s) \simeq \frac{1}{C_{\alpha}  \mathcal{D}_{\alpha} ^{\frac{1}{2+\alpha}} s^{\frac{1+\alpha}{2+\alpha}}}-\frac{2M}{C_{\alpha}^2  \mathcal{D}_{\alpha} ^{\frac{2}{2+\alpha}} s^{\frac{\alpha}{2+\alpha}}}.
\label{appen-asy-PM-eq-5}
\end{align}
Finally, we perform the inverse Laplace tramsformation and obtain
\begin{align}
P_m(M|t) = \frac{1}{\left( \mathcal{D}_{\alpha} t\right)^{\frac{1}{2+\alpha}}} \mathcal{F}_{\alpha} \left( \frac{M}{\left( \mathcal{D}_{\alpha} t\right)^{\frac{1}{2+\alpha}}} \right),
\label{appen-asy-PM-eq-6}
\end{align}
where the scaling function $\mathcal{F}_{\alpha} (z )$ has the form
\begin{align}
\mathcal{F}_{\alpha} (z \to 0) & \simeq \frac{1}{C_{\alpha} \Gamma \left( \frac{1+\alpha}{2+\alpha}\right)}-\frac{2z}{C_{\alpha} ^2 \Gamma\left( \frac{\alpha}{2+\alpha}\right)},~~~\text{for }\alpha \neq 0.
\label{appen-asy-PM-eq-7}
\end{align}
On the other hand, for $\alpha = 0 $, we saw that $\mathcal{F}_{\alpha}(z) = \frac{e^{-z^2/4}}{\sqrt{\pi}}$ [see Eq. (31) in the main text] which for small $z$ becomes
\begin{align}
\mathcal{F}_{\alpha} (z \to 0) \simeq \frac{1}{\sqrt{\pi}} \left( 1-\frac{z^2}{4}\right),~~~~\text{for }\alpha = 0.
\label{appen-asy-PM-eq-8}
\end{align}
Eqs. \eqref{appen-asy-PM-eq-7} and \eqref{appen-asy-PM-eq-8} completely specifies the small-$z$ behaviour of the scaling function $\mathcal{F}_{\alpha}(z) $ for all values of $\alpha$.

\subsection{Large $z$ behaviour of $\mathcal{F}_{\alpha}(z)$}
We next look at the large-$z$ behaviour of $\mathcal{F}_{\alpha}(z)$. Once again, this comes down to analysing the large-$M$ behaviour of $\bar{J}(M,s)$ via Eq. \eqref{new-appen-extreme-eq-15}. Using the asmptotic expression of modified Bessel function for large $x$ as
\begin{align}
I_{\beta}(x) \simeq \frac{e^{x}}{\sqrt{2 \pi x}},
\label{appen-asy-PM-eq-9}
\end{align}
in Eq. \eqref{new-appen-extreme-eq-8}, we get $\mathcal{H}_{\beta}(x)$ as 
\begin{align}
\mathcal{H}_{\beta}(x) \simeq \sqrt{\frac{2}{\pi}} \frac{e^x}{x^{\frac{1}{2}-\beta}},~~~~~ \text{as }x \to \infty.
\label{appen-asy-PM-eq-10}
\end{align}
Substituting this in the expression of $\bar{J}(M,s)$ in Eq. \eqref{new-appen-extreme-eq-16}, we find 
\begin{align}
\bar{J}(M,s) \simeq \frac{\mathcal{H}_{\frac{1}{2+\alpha}}(0)}{s} \sqrt{\frac{\pi}{2}} \left( a_s M\right)^{\frac{\alpha}{4}} e^{-\left( a_s M\right)^{\frac{2+\alpha}{2}}}.
\label{appen-asy-PM-eq-11}
\end{align}
as $M \to \infty$. Inserting this in the expression of $\bar{P}_m(M|s)$ in Eq. \eqref{new-appen-extreme-eq-15} and performing some algebraic simplications, we get
\begin{align}
\bar{P}_m(M|s) \simeq \frac{Z_{\alpha}~\text{exp} \left( -2b_M \sqrt{s} \right)}{s^{\frac{4+\alpha}{4(2+\alpha)}}} ,~~\text{as }M \to \infty,
\label{appen-asy-PM-eq-12}
\end{align}
where $b_M$ and $Z_{\alpha}$ are defined as
\begin{align}
& b_M = \frac{1}{2}\sqrt{\frac{M^{2+\alpha}}{\mathcal{D}_{\alpha}}}, \label{appen-asy-PM-eq-13} \\
& Z_{\alpha} = \frac{(2+\alpha)\sqrt{\pi}}{2\sqrt{2}} \left(  \frac{M^{\frac{3 \alpha}{4}}\mathcal{H}_{\frac{1}{2+\alpha}}(0)}{\mathcal{D}_{\alpha}^{\frac{4+3\alpha}{4(2+\alpha)}}}\right).\label{appen-asy-PM-eq-14}
\end{align}
We now proceed to perform the inverse Laplace transformation of $\bar{P}_m(M|s)$ in Eq. \eqref{appen-asy-PM-eq-12}. To this end, we exploit the convolution property of Laplace transformation to write
\begin{align}
P_m(M|t) \simeq &  ~Z_{\alpha} \int _{0}^{t} dT ~\mathcal{L}_{s \to t- T} \left[\frac{1}{s^{\frac{4+\alpha}{4(2+\alpha)}}} \right] ~~~~~~~~~~~~~~~~~~~~~~~~\nonumber \\
& ~~~~~~~~~ \times \mathcal{L}_{s \to T} \left[ \text{exp} \left( -2b_M \sqrt{s} \right) \right],
\label{appen-asy-PM-eq-15}
\end{align} 
where the notation $\mathcal{L}_{s \to t}$ denotes the inverse Laplace transformation. Using the inverse Laplace transformations
\begin{align}
&\mathcal{L}_{s \to t}\left[ \text{exp} \left( -2b_M \sqrt{s} \right) \right] = \frac{b_M~e^{-\frac{b_m^2}{t}}}{\sqrt{\pi t^3}},
\label{appen-asy-PM-eq-16} \\
&\mathcal{L}_{s \to t} \left[ s^{-\frac{4+\alpha}{4(2+\alpha)}}\right] = \frac{t^{\frac{4+\alpha}{4(2+\alpha)}-1}}{\Gamma\left( \frac{4+\alpha}{4(2+\alpha)} \right) }, \label{appen-asy-PM-eq-17}
\end{align}
in Eq. \eqref{appen-asy-PM-eq-15}, we find
\begin{align}
P_m(M|t) \simeq \frac{Z_{\alpha} M^{\frac{2+\alpha}{2}}~\mathbb{I} \left( \frac{M^{2+\alpha}}{4 \mathcal{D}_{\alpha} t} \right)}{\sqrt{4 \pi \mathcal{D}_{\alpha}} ~\Gamma \left(\frac{4+\alpha}{8+4 \alpha} \right) t^{\frac{8+5\alpha}{8+4 \alpha}}}, 
\label{appen-asy-PM-eq-18}
\end{align} 
where the function $\mathbb{I}(y)$ is defined as
\begin{align}
\mathbb{I}(y) = \int _{0}^{1} dw \frac{e^{-y/w}}{w^{3/2} (1-w)^{\frac{4+3 \alpha}{8+4 \alpha}}},~~~~y>0.
\label{appen-asy-PM-eq-19}
\end{align}
As seen in Eq. \eqref{appen-asy-PM-eq-18}, we now need to evaluate $\mathbb{I} \left( \frac{M^{2+\alpha}}{4 \mathcal{D}_{\alpha} t} \right)$ for large $M$ in order to compute the distribution $P_m(M|t)$. In Sec. \ref{appen-IIa}, we have shown that the asymptotic expression of the function $\mathbb{I}(y)$ is
\begin{align}
\mathbb{I}(y) \simeq \frac{e^{-y}}{y^{\frac{4+\alpha}{8+4 \alpha}}}~\Gamma \left(\frac{4+\alpha}{8+4 \alpha} \right),~~~\text{as }y \to \infty.
\label{appen-asy-PM-eq-20}
\end{align}
Finally, plugging this in Eq. \eqref{appen-asy-PM-eq-18}, we find
\begin{align}
P_m(M|t) = \frac{1}{\left( \mathcal{D}_{\alpha} t\right)^{\frac{1}{2+\alpha}}} \mathcal{F}_{\alpha} \left( \frac{M}{\left( \mathcal{D}_{\alpha} t\right)^{\frac{1}{2+\alpha}}} \right),
\label{appen-asy-PM-eq-21}
\end{align}
with the scaling function given by
\begin{align}
\mathcal{F}_{\alpha} \left( z \to \infty \right)\simeq \frac{(2+\alpha)\mathcal{H}_{\frac{1}{2+\alpha}}(0) }{2^{\frac{3+2\alpha}{2+\alpha}}}  z^{\alpha} ~e^{-\frac{z^{2+\alpha}}{4} },
\label{appen-asy-PM-eq-22}
\end{align}
as $z \to \infty$. This expression has been quoted in Eq. (33) in the main text.

\subsection{$\mathbb{I}(y)$ in Eq. \eqref{appen-asy-PM-eq-19} as $y \to \infty$}
\label{appen-IIa}
Here, we evaluate the asymptotic form of the integral $\mathbb{I}(y)$ in Eq. \eqref{appen-asy-PM-eq-20} as $y \to \infty$ which was instrumental in computing $P_m(M|t)$ for large $M$. For this, we  first replace $w=\bar{w}^{-1}$ in Eq. \eqref{appen-asy-PM-eq-19} and rewrite
\begin{align}
\mathbb{I}(y) = \int _{1}^{\infty} \frac{d \bar{w}}{\bar{w}^{\frac{\alpha}{8+4 \alpha}}}~\frac{e^{-y \bar{w}}}{\left( \bar{w}-1 \right)^{\frac{4+3 \alpha}{8+4 \alpha}}}.
\label{appen-asy-PM-eq-22}
\end{align}
For $y \to \infty$, the integral will be dominated by small values of $\bar{w}$ which in the given domain of integration is equal to $1$. Consequently, we get
\begin{align}
\mathbb{I} \left( y \to \infty \right) \simeq \int _{1}^{\infty} d \bar{w}~\frac{e^{-y \bar{w}}}{\left( \bar{w}-1 \right)^{\frac{4+3 \alpha}{8+4 \alpha}}}.
\label{appen-asy-PM-eq-23}
\end{align}
Finally, changing the variable $\bar{w}=w+1 $, we get
\begin{align}
\mathbb{I} \left( y \to \infty \right) &\simeq e^{-y}~\int _{0}^{\infty} d w~\frac{e^{-y w}}{w^{\frac{4+3 \alpha}{8+4 \alpha}}}, 
\\
& \simeq \frac{e^{-y}}{y^{\frac{4+\alpha}{8+4 \alpha}}}~\Gamma \left(\frac{4+\alpha}{8+4 \alpha} \right).
\label{appen-asy-PM-eq-23}
\end{align}
This result has been quoted in Eq. \eqref{appen-asy-PM-eq-20}.

\begin{figure}[t]
\includegraphics[scale=0.4]{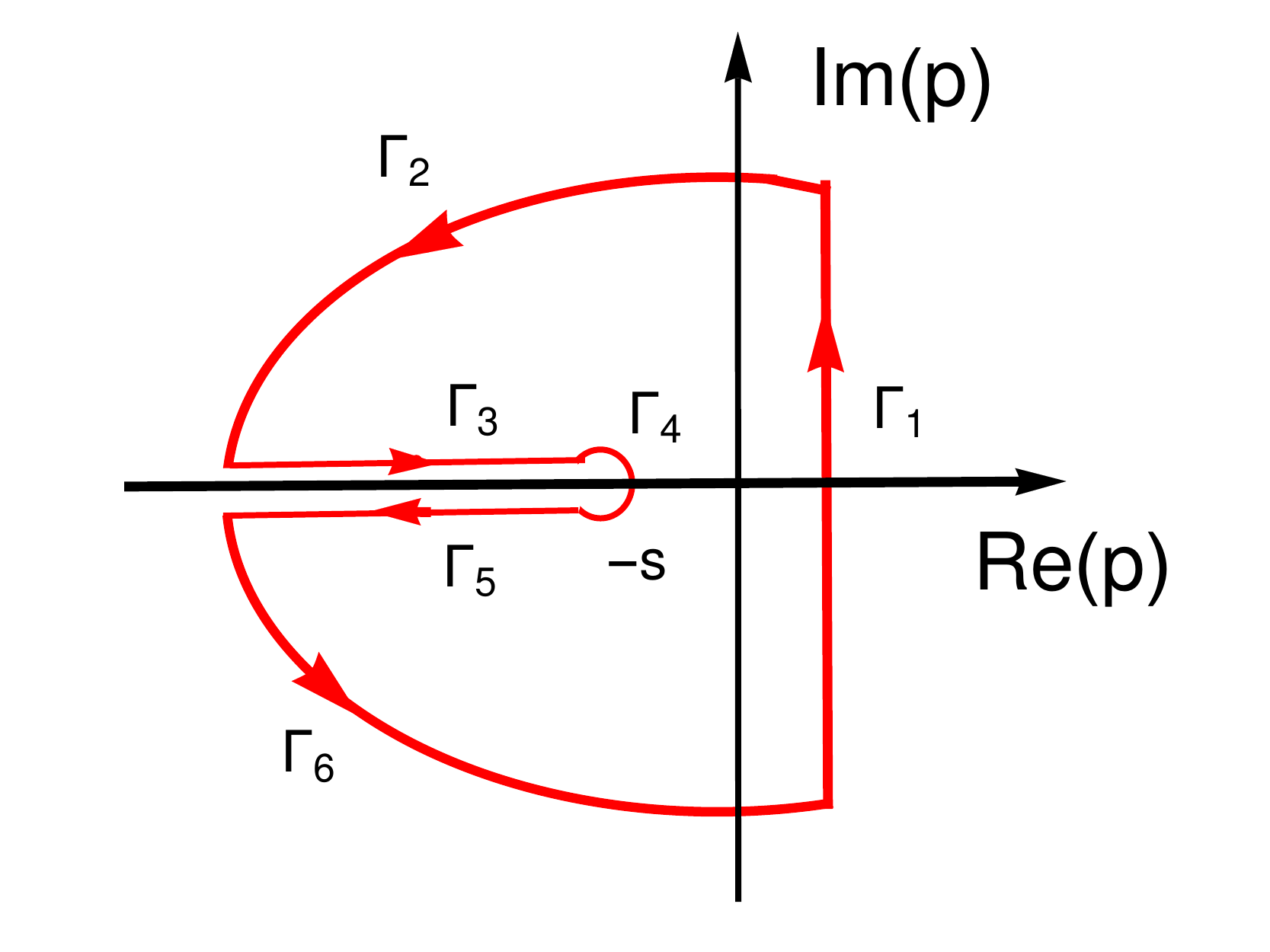}
\centering
\caption{Contour for the Bromwhich integration of $\mathbb{E}(s,t_m,w) $ in Eq. \eqref{appen-ILTYY-eq-4}}
\label{contour-fig-2}
\end{figure}
\section{Marginal distribution $\mathcal{P}_m(t_m|t)$}
\label{ILTYY}
Let us now look at the distribution $\mathcal{P}_m(t_m|t)$ of the arg-maximum $t_m$. In Eq. (36) of the main text, we had obtained the double Laplace transform $\bar{\mathcal{P}}_m(p|s)$ as
\begin{align}
\bar{\mathcal{P}}_m(p|s) &= \int _{0}^{\infty} dM \frac{  \mathcal{H}_{\frac{1}{2+\alpha}} \left(0 \right) }{s~\mathcal{H}_{\frac{1}{2+\alpha}} \left( (a_{s+p} M)^{\frac{2+\alpha}{2}} \right)}
 \frac{\partial _M \left[ \mathcal{H}_{\frac{1}{2+\alpha}} \left( (a_s M)^{\frac{2+\alpha}{2}} \right) \right]}{ \mathcal{H}_{\frac{1}{2+\alpha}} \left( (a_s M)^{\frac{2+\alpha}{2}} \right)}.
\label{new-appen-extreme-eq-30}
\end{align}
We further simplify this expression by using the identity
$$\frac{d}{dy}  \mathcal{H}_{\frac{1}{2+\alpha}} \left( y^{\frac{2+\alpha}{2}} \right) = \left(\frac{2+\alpha}{2}\right) \mathcal{H}_{\frac{1+\alpha}{2+\alpha}} \left( y^{\frac{2+\alpha}{2}} \right), $$
from Eq. \eqref{new-appen-extreme-eq-8} and change the variable $a_{s+p} M=w$. Eq. \eqref{new-appen-extreme-eq-30} can then be rewritten as
\begin{align}
&\bar{\mathcal{P}}_m(p|s) = \frac{ \mathcal{H}_{\frac{1}{2+\alpha}} \left(0 \right) }{2 (2+\alpha)^{-1}} \int _{0}^{\infty} dw \frac{\bar{\mathbb{Y}}_{\alpha} \left(s,p,w \right)}{\mathcal{H}_{\frac{1}{2+\alpha}} \left(w^{\frac{2+\alpha}{2}} \right)}, ~~\text{with}\label{extreme-eq-31}\\
& \bar{\mathbb{Y}}_{\alpha} \left(s,p,w \right) = \frac{\mathcal{H}_{\frac{1+\alpha}{2+\alpha}} \left(\sqrt{\frac{s}{s+p}}w^{\frac{2+\alpha}{2}} \right)}{s^{\frac{1+\alpha}{2+\alpha}}(s+p)^{\frac{1}{2+\alpha}}~\mathcal{H}_{\frac{1}{2+\alpha}} \left(\sqrt{\frac{s}{s+p}}w^{\frac{2+\alpha}{2}} \right)}.\label{extreme-eq-32}
\end{align} 
Notice that all $s$ and $p$ dependences in Eq. \eqref{extreme-eq-31} now appear in the function $\bar{\mathbb{Y}}_{\alpha} \left(s,p,w \right) $. To get the distribution in time domain, one then needs to perform the double inverse Laplace transformation of $\bar{\mathbb{Y}}_{\alpha} \left(s,p,w \right)$. Denoting this double inverse Laplace transformation by $\mathbb{Y}_{\alpha} \left(t,t_m,w \right)$, \textit{i.e.} 
\begin{align}
\mathbb{Y}_{\alpha} \left(t,t_m,w \right) = \mathcal{L}_{s \to t}\mathcal{L}_{p \to t_m} \left[ \bar{\mathbb{Y}}_{\alpha} \left(s,p,w \right) \right].
\label{appen-ILTYY-eq-2}
\end{align}
Notation $\mathcal{L}_{s \to t}$ denotes the inverse Laplace transformation from $s \to t$. Using Eq. \eqref{appen-ILTYY-eq-2}, we can now write the distribution $\mathcal{P}_m(t_m|t)$ as
\begin{align}
\mathcal{P}_m(t_m|t) = \frac{ \mathcal{H}_{\frac{1}{2+\alpha}} \left(0 \right) }{2 (2+\alpha)^{-1}} \int _{0}^{\infty} dw \frac{\mathbb{Y}_{\alpha} \left(t,t_m,w \right)}{\mathcal{H}_{\frac{1}{2+\alpha}} \left(w^{\frac{2+\alpha}{2}} \right)}.
\label{appen-ILTYY-eq-1}
\end{align}  
Let us now proceed to perform the inversion in Eq. \eqref{appen-ILTYY-eq-2}. For the inversion from $p \to t_m$, we write the Bromwich integral
\begin{align}
\mathbb{E}(s,t_m,w) &= \mathcal{L}_{p \to t_m} \left[ \bar{\mathbb{Y}}_{\alpha} \left(s,p,w \right) \right], \label{appen-ILTYY-eq-3}\\
& =\frac{1}{2 \pi i} \int _{-i \infty}^{i \infty} dp~ e^{p t_m}~ \bar{\mathbb{Y}}_{\alpha} \left(s,p,w \right) \label{appen-ILTYY-eq-4}.
\end{align}
Note that the integrand $\bar{\mathbb{Y}}_{\alpha} \left(s,p,w \right)$ has a branch point at $p=-s$ [see Eq. \eqref{extreme-eq-32}]. To perform this complex integration, we consider contour shown in Figure \ref{contour-fig-2}. Since the integrand $\bar{\mathbb{Y}}_{\alpha} \left(s,p,w \right)$ is analytic inside this contour, the Cauchy theorem gives
\begin{align}
\int _{\Gamma _1}+\int _{\Gamma _2}+\int _{\Gamma _3}+\int _{\Gamma _4}+\int _{\Gamma _5}+\int _{\Gamma _6} =0. 
\label{appen-ILTYY-eq-5}
\end{align}
Here $\int _{\Gamma _1}$ is the integration that we need. Let us now calculate these integrations along different paths. Note that the real part of $p$ is negative along paths $\Gamma_2$ and $\Gamma _6$ and in the limit $|p| \to \infty$, the integrand becomes exactly zero. Hence, the contribution of these two paths is equal to zero.

Along $\Gamma _4$, we replace $p=-s + \delta e^{i \theta}$ and take limit $\delta \to 0^+$. For small $\delta$, we find that the integrand scales as $\sim \sqrt{\delta}$ which vanishes as $\delta \to 0^+$. Hence, the contribution of this path is also zero. 

Next, we look at the contribution of $\Gamma _3$. For this path, we replace $p=-s+u s e^{i \pi} w^{2+\alpha} $ and perform some algebraic simplifications to get
\begin{align}
\int _{\Gamma _3} = \frac{w^{1+\alpha}~e^{-\frac{i \pi }{2+\alpha}}}{2 \pi i} \int _{0}^{\infty} du \frac{e^{-(1+u w^{2+\alpha}) s t_m}}{u^{\frac{1}{2+\alpha}}} \frac{\mathcal{H}_{\frac{1+\alpha}{2+\alpha}} \left( -\frac{i}{\sqrt{u}}\right)}{\mathcal{H}_{\frac{1}{2+\alpha}} \left( -\frac{i}{\sqrt{u}}\right)}. \nonumber
\end{align}  
Similarly, for $\Gamma _5$, we replace $p=-s+u s e^{-i \pi} w^{2+\alpha} $ and get 
 \begin{align}
\int _{\Gamma _5} = -\frac{w^{1+\alpha}~e^{\frac{i \pi }{2+\alpha}}}{2 \pi i} \int _{0}^{\infty} du \frac{e^{-(1+u w^{2+\alpha}) s t_m}}{u^{\frac{1}{2+\alpha}}} \frac{\mathcal{H}_{\frac{1+\alpha}{2+\alpha}} \left( \frac{i}{\sqrt{u}}\right)}{\mathcal{H}_{\frac{1}{2+\alpha}} \left( \frac{i}{\sqrt{u}}\right)}. \nonumber
\end{align}  
Plugging all these contribution in Eq. \eqref{appen-ILTYY-eq-5} and noting $\int _{\Gamma _1} = \mathbb{E}(s,t_m,w)$ , we find
\begin{align}
\mathbb{E}(s,t_m,w) = \int _{0}^{\infty} du~ \frac{w^{1+\alpha}~e^{-(1+u w^{2+\alpha}) s t_m}}{u^{\frac{1}{2+\alpha}}}~\mathbb{X}_{\alpha} \left( \sqrt{u}\right), 
\label{appen-ILTYY-eq-6}
\end{align}
where the function $\mathbb{X}_{\alpha} \left( u\right)$ is defined as
\begin{align}
\mathbb{X}_{\alpha} \left( u\right) =\frac{e^{\frac{i \pi }{2+\alpha}}}{2 \pi i} \left[\frac{\mathcal{H}_{\frac{1+\alpha}{2+\alpha}} \left( \frac{i}{{u}}\right)}{\mathcal{H}_{\frac{1}{2+\alpha}} \left( \frac{i}{{u}}\right)} -\frac{e^{-\frac{2i \pi }{2+\alpha}}~\mathcal{H}_{\frac{1+\alpha}{2+\alpha}} \left( -\frac{i}{{u}}\right)}{\mathcal{H}_{\frac{1}{2+\alpha}} \left(- \frac{i}{{u}}\right)} \right].
\label{new-appen-ILTYY-eq-7}
\end{align}
Inserting $\mathbb{E}(s,t_m,w)$ from Eq. \eqref{appen-ILTYY-eq-6} in $\mathbb{Y}_{\alpha} \left(t,t_m,w \right)$ in Eq. \eqref{appen-ILTYY-eq-2} and performing the inversion with $s$, we get
\begin{align}
\mathbb{Y}_{\alpha} \left(t,t_m,w \right) = \frac{\mathbb{X}_{\alpha} \left( \sqrt{\frac{t-t_m}{t_m w^{2+\alpha}}}  \right)}{t_m^{\frac{1+\alpha}{2+\alpha}} (t-t_m)^{\frac{1}{2+\alpha}}}.
\label{appen-ILTYY-eq-71}
\end{align}
Finally, substituting this form of $\mathbb{Y}_{\alpha} \left(t,t_m,w \right)$ in Eq. \eqref{appen-ILTYY-eq-1} results in the expression of $\mathcal{P}_m(t_m|t)$ as
\begin{align}
\mathcal{P}_m(t_m|t) = \frac{1}{t} ~\mathcal{G} _{m}^{\alpha} \left( \frac{t_m}{t}\right),
\label{new-appen-extreme-res-eq-1}
\end{align}
with the scaling function $\mathcal{G} _{m}^{\alpha} \left( z \right)$ defined as
\begin{align}
~~~~~~~~\mathcal{G} _{m}^{\alpha} \left( z \right) = \frac{(2+\alpha)\mathcal{H}_{\frac{1}{2+\alpha}} \left(0 \right)}{2z^{\frac{1+\alpha}{2+\alpha}} (1-z)^{\frac{1}{2+\alpha}}}~\int _{0}^{\infty} dw\frac{\mathbb{X}_{\alpha} \left( \sqrt{\frac{1-z}{z w^{2+\alpha}}} \right)}{\mathcal{H}_{\frac{1}{2+\alpha}}\left( w^{\frac{2+\alpha}{2}}\right)}.
\label{new-appen-extreme-res-eq-2}
\end{align}

\section{Asymptotic form of $\mathcal{G}_m^{\alpha}(z)$}
\label{apen-Gm}
We next analyse the asymptotic behaviour of the scaling function $\mathcal{G}_m^{\alpha}(z)$ in Eq. \eqref{new-appen-extreme-res-eq-2} as $z \to 0$ and $z \to 1$. For clarity, we present this analysis separately for $z \to 0$ and $z \to 1$.
\subsection{$\mathcal{G}_m^{\alpha}(z)$ as $z \to 0$}
To get the asymptotic form of $\mathcal{G}_m^{\alpha}(z)$ as $z \to 0$, we see from Eq. \eqref{new-appen-extreme-res-eq-2} that one needs to specify the behaviour of $\mathbb{X}_{\alpha} \left( \sqrt{\frac{1-z}{z w^{2+\alpha}}} \right)$ as $z \to 0$. To this end, we use the expression of $\mathbb{X}_{\alpha}(x)$ in Eq. \eqref{new-appen-ILTYY-eq-7} as $x \to \infty$ which reads
\begin{align}
\mathbb{X}_{\alpha}\left(x \to \infty \right) \simeq \frac{\sin \left( \frac{\pi}{2+\alpha}\right)}{ \pi} \frac{\mathcal{H}_{\frac{1+\alpha}{2+\alpha}}(0)}{\mathcal{H}_{\frac{1}{2+\alpha}}(0)}.
\end{align}  
Plugging this in Eq. \eqref{new-appen-extreme-res-eq-2}, we see that
\begin{align}
\mathcal{G}_m^{\alpha}(z) \sim z^{-\frac{1+\alpha}{2+\alpha}}, ~~~~\text{as } z \to 0.
\end{align}
\subsection{$\mathcal{G}_m^{\alpha}(z)$ as $z \to 1$}
Once again, we use the expression $\mathcal{G}_m^{\alpha}(z)$ in Eq. \eqref{new-appen-extreme-res-eq-2}. As evident from this equation, one then needs to specify the small-$x$ behaviour of $\mathbb{X}_{\alpha}(x)$. Using $\mathcal{H}_{\beta}\left(u \to \infty\right) \simeq \sqrt{\frac{1}{\pi}} e^{u} u^{\beta -\frac{1}{2}}$ in Eq. \eqref{new-appen-ILTYY-eq-7}, it is easy to show that
\begin{align}
\mathbb{X}_{\alpha}\left(x \to 0 \right) \simeq \frac{1}{\pi x^{\frac{\alpha}{2+\alpha}}}.
\end{align} 
Inserting this in Eq. \eqref{new-appen-extreme-res-eq-2} then yields
\begin{align}
\mathcal{G}_m^{\alpha}(z) \sim (1-z)^{-\frac{1}{2}},~~~\text{as } z \to 1.
\end{align}

\section{Residence time distribution $\mathcal{P}_r \left( t_r|t \right)$}
\label{appen-sol-resi}
We now look at the second arcsine law which concerns the statistics of the residence time $t_r$. We saw that the Laplace transform  $\mathcal{Q}(p,x_0|t)$ for the residence time obeys the backward master equation
\begin{align}
\partial _t \mathcal{Q}(p,x_0|t) = \left[D(x_0) \partial _{x_0}^2-p~ \Theta (x_0) \right]\mathcal{Q}(p,x_0|t),
\label{new-appen-resi-eq-3}
\end{align}
where $D(x_0)$ is given in Eq. \eqref{new-appen-extreme-eq-3}. The corresponding initial and boundary conditions were derived in Sec.  IV in the main text. They read
\begin{align}
& \mathcal{Q}(p,x_0|t \to 0) = 1, \label{new-appen-resi-eq-4}\\
& \mathcal{Q}(p,x_0 \to -\infty |t) = 1, \label{new-appen-resi-eq-5} \\
& \mathcal{Q}(p,x_0 \to \infty |t) = e^{-pt}. \label{new-appen-resi-eq-6}
\end{align}

Here we will solve the backward master equation \eqref{new-appen-resi-eq-3} along with these conditions to get the probability distribution of the residence time $t_r(t)$. For this, we first take the Laplace transformation of Eq. \eqref{new-appen-resi-eq-3} with respect to $t$ and rewrite it as
\begin{align}
\left[ s+p ~\Theta (x_0) \right] \bar{\mathcal{Q}}(p,x_0|s) - 1= D(x_0) \partial _{x_0}^2\bar{\mathcal{Q}}(p,x_0|s),
\label{appen-sol-resi-eq-1}
\end{align}
where $\bar{\mathcal{Q}}(p,x_0|s)$ is the Laplace transformation of $\mathcal{Q}(p,x_0|t)$. For $x_0>0$, this equation becomes
\begin{align}
 (s+p) \bar{\mathcal{Q}}(p,x_0|s) - 1= D(x_0) \partial _{x_0}^2\bar{\mathcal{Q}}(p,x_0|s). 
\label{appen-sol-resi-eq-2}
\end{align}
To simplify this equation further, we make the following transformations: 
\begin{align}
&y = \frac{x_0 ^{2+\alpha}}{(2+\alpha)^2 l^{\alpha} D_0}, \\
&\bar{\mathcal{Q}}(p,x_0|s) = \frac{1}{s+p}+\mathbb{Q}(p,x_0|s),
\end{align}
and rewrite Eq. \eqref{appen-sol-resi-eq-2} in terms of $\mathbb{Q}(p,x_0|s)$ and $y$ as
\begin{align}
y \frac{\partial ^2 \mathbb{Q}}{\partial y^2} + \left(\frac{1+\alpha}{2+\alpha} \right) \frac{\partial  \mathbb{Q}}{\partial y} = (s+p) \mathbb{Q}.
\label{appen-sol-resi-eq-4}
\end{align}
This equation can now be solved and its solutions are given in terms of the modified bessel functions as $y^{\frac{1}{2(2+\alpha)}} I_{\frac{1}{2+\alpha}} \left( 2 \sqrt{(s+p) y}\right)$ and $y^{\frac{1}{2(2+\alpha)}} K_{\frac{1}{2+\alpha}} \left( 2 \sqrt{(s+p) y}\right)$. However, the former solution diverges in the limit $y \to \infty$. Hence, we consider only the later solution and write finally for $\bar{\mathcal{Q}}(p,x_0|s)$ as
\begin{align}
\bar{\mathcal{Q}}(p,x_0|s) & = \frac{1}{s+p} +\mathbb{C}_5 \sqrt{x_0}~K_{\frac{1}{2+\alpha}} \left( (a_{s+p}~ x_0)^{\frac{2+\alpha}{2}}\right),
\label{appen-sol-resi-eq-5}
\end{align}
where $a_{s+p}$ is given in Eq. \eqref{new-appen-extreme-eq-9} and $\mathbb{C}_5$ is a function independent of $y$, but can, in principle, depend on $s$ and $p$. Recall that the solution in Eq. \eqref{appen-sol-resi-eq-5} holds only for $x_0 >0$. Proceeding similarly for $x_0 <0$, we get
\begin{align}
\bar{\mathcal{Q}}(p,x_0|s) & = \frac{1}{s} +\mathbb{C}_6 \sqrt{|x_0|}~K_{\frac{1}{2+\alpha}} \left( (a_{s}~| x_0|)^{\frac{2+\alpha}{2}}\right).
\label{appen-sol-resi-eq-61}
\end{align}
Once again $\mathbb{C}_6$ here is a function independent of $y$, but can, in principle, depend on $s$ and $p$. Now the task is to compute these functions $\mathbb{C}_5$ and $\mathbb{C}_6$ in Eqs. \eqref{appen-sol-resi-eq-5} and \eqref{appen-sol-resi-eq-61}. For this computation, we use the continuity of $\bar{\mathcal{Q}}(p,x_0|s) $ and $\partial _{x_0} \bar{\mathcal{Q}}(p,x_0|s) $ across $x_0 = 0$:
\begin{align}
\bar{\mathcal{Q}} \left(p,x_0 \to 0^+|s \right)& =\bar{\mathcal{Q}} \left(p,x_0 \to 0^-|s \right), \\
\left( \frac{\partial \bar{\mathcal{Q}} \left(p,x_0|s \right)}{\partial x_0}\right) _{x_0 \to 0^+}& = \left( \frac{\partial \bar{\mathcal{Q}} \left(p,x_0|s \right)}{\partial x_0}\right) _{x_0 \to 0^-}. \label{pras-neww} 
\end{align}
Then using these forms of $\mathbb{C}_5$ and $\mathbb{C}_6$ in Eqs. \eqref{appen-sol-resi-eq-5} and \eqref{appen-sol-resi-eq-61}, we obtain the expression of $\bar{\mathcal{Q}}(p,x_0|s)$ for all $x_0$. However, we are interested only in the situation where the particle starts initially from the origin. For this case, we have
\begin{align}
\bar{\mathcal{Q}}(p|s) = \frac{1}{s}-\frac{p}{s(s+p)} \left[1+ \left( \frac{s}{s+p} \right) ^{\frac{1}{2+\alpha}} \right]^{-1},
\label{appen-sol-resi-eq-6}
\end{align}
where we have used the short hand notation $\bar{\mathcal{Q}}(p|s)$ for $\bar{\mathcal{Q}}(p,x_0=0|s)$. 
Performing inversion of the Laplace transform in Eq. \eqref{appen-sol-resi-eq-6} for $\alpha >-1$, we get \cite{newLamperti58, newCarmi2010}
\begin{align}
\mathcal{P}_r(t_r|t) = \frac{1}{t} ~\mathcal{G} _{r}^{\alpha} \left( \frac{t_r}{t}\right),
\label{new-appen-resi-eq-8}
\end{align}
where the scaling function $\mathcal{G} _{r}^{\alpha} \left( z \right)$ is given by
\begin{align}
& ~~~~~~~~~~\mathcal{G} _{r}^{\alpha} \left( z \right) = \frac{ \sin \left( \frac{\pi}{2+\alpha}\right)  \pi^{-1}~ \left[ z(1-z) \right]^{-\frac{1+\alpha}{2+\alpha}}}{z^{\frac{2}{2+\alpha}}+(1-z)^{\frac{2}{2+\alpha}}+2  \cos \left( \frac{\pi}{2+\alpha}\right) \left[z(1-z) \right] ^{\frac{1}{2+\alpha}}}.
\label{new-appen-resi-eq-9}
\end{align}

\section{Distribution $\mathcal{P}_{ \ell } \left( t_{\ell}|t \right)$ of the last-passage time}
\label{appen-dist-Pl}
This section deals with the arcsine law for the last-passage time $t_{\ell}$. As discussed in Sec. IV in the main text, the Laplace transform $\bar{\mathcal{P}}_{ \ell } \left( p|s \right)$ of this distribution reads
\begin{align}
\bar{\mathcal{P}}_{ \ell } \left( p|s \right) = \frac{\bar{\mathbb{P}}(\epsilon, s+p|0)~\bar{S}_0(s|\epsilon)}{\mathcal{N}_L(\epsilon)}. 
\label{new-appen-last-passage-time-eq-2}
\end{align}
where $\bar{\mathbb{P}}(x, s|0)$ stands for the Laplace transformation of the distribution $\mathbb{P}(x,t|0)$ in the free space. Also, $\bar{S}_0(s|x)$ is the Laplace transform of the survival probability as discussed before. In what follows, we first compute these two Laplace transforms and then use them in Eq. \eqref{new-appen-last-passage-time-eq-2} to calculate $\bar{\mathcal{P}}_{ \ell } \left( p|s \right)$.
\subsection{Computation of $\bar{\mathbb{P}}(\epsilon, s|0)$}
\label{appen-dist-Pl-prob}
Denoting the probability distribution to be at $x$ at time $t$ starting from the origin by $\mathbb{P}(x,t|0)$, we write the Fokker-Planck equation (Ito sense) for $\mathbb{P}(x,t|0)$ as 
\begin{align}
\partial _t \mathbb{P}(x,t|0) = \partial _{x}^2 \left[ D(x) \mathbb{P}(x,t|0) \right],
\label{appen-dist-Pl-eq-1}
\end{align} 
where $D(x)$ is given in Eq. \eqref{new-appen-extreme-eq-3}. Taking Laplace tramsformation of this equation gives  
\begin{align}
s \bar{\mathbb{P}}(x,s|0)-\delta(x) = \partial _{x}^2 \left[ D(x) \bar{\mathbb{P}}(x,s|0) \right].
\label{appen-dist-Pl-eq-2}
\end{align}
For $x \neq 0$, we get rid of the delta function and obtain
\begin{align}
s \bar{\mathbb{P}}(x,s|0) = \partial _{x}^2 \left[ D(x) \bar{\mathbb{P}}(x,s|0) \right].
\label{appen-dist-Pl-eq-3}
\end{align}
Let us first analyse this differential equation for $x>0$. To simplify Eq. \eqref{appen-dist-Pl-eq-3}, we perform the transformations
\begin{align}
 & y = \frac{x^{2+\alpha}}{(2+\alpha)^2 l^{\alpha} D_0}, \label{appen-dist-Pl-eq-4} \\
 & \mathbb{Z}_L \left(x,s \right)= D(x) \bar{\mathbb{P}}(x,s|0), \label{appen-dist-Pl-eq-5} 
\end{align}
and rewrite it as
\begin{align}
y \frac{\partial ^2 \mathbb{Z}_L}{\partial y^2} + \left(\frac{1+\alpha}{2+\alpha} \right) \frac{\partial  \mathbb{Z}_L}{\partial y} = s \mathbb{Z}_L.
\label{appen-dist-Pl-eq-6} 
\end{align}
One can now solve this equation exactly and write solutions in terms of the modified bessel functions as $y^{\frac{1}{2(2+\alpha)}} I_{\frac{1}{2+\alpha}} \left( 2 \sqrt{s y}\right)$ and $y^{\frac{1}{2(2+\alpha)}} K_{\frac{1}{2+\alpha}} \left( 2 \sqrt{s y}\right)$. Note that the former solution diverges in the limit $y \to \infty$. Hence, we consider only the later solution and write finally $\bar{\mathbb{P}}(x,s|0)$ as
\begin{align}
\bar{\mathbb{P}}(x,s|0) & = \frac{\mathbb{C}_7 \sqrt{x}}{D(x)}~K_{\frac{1}{2+\alpha}} \left( (a_{s}~ x)^{\frac{2+\alpha}{2}}\right),,
\label{appen-dist-Pl-eq-7} 
\end{align}
where $a_{s}$ is given in Eq. \eqref{new-appen-extreme-eq-9} and $\mathbb{C}_7$ is a constant which, in principle, depends of $s$ and $\alpha$. Recall that the expression of $\bar{\mathbb{P}}(x,s|0)$ in Eq. \eqref{appen-dist-Pl-eq-7} is valid only for $x>0$. To compute this for $x<0$, we recall that the problem possesses $x \to -x$ symmetry in the infinite line. Consequently, the same solution also applies for $x<0$ and we have
\begin{align}
\bar{\mathbb{P}}(x,s|0) & = \frac{\mathbb{C}_7 \sqrt{x}}{D(x)}~K_{\frac{1}{2+\alpha}} \left( (a_{s}~ |x|)^{\frac{2+\alpha}{2}}\right),
\label{appen-dist-Pl-eq-81} 
\end{align}
for all $x$. The task now is to compute the constant $\mathbb{C}_7$. To compute it, we integrate Eq. \eqref{appen-dist-Pl-eq-2} with respect to $x$ from $\zeta$ to $-\zeta$ and take $\zeta \to 0^+$. This gives rise to the following condition:
\begin{align}
\left( \frac{ \partial\left[ D(x) \bar{\mathbb{P}}(x,s|0) \right] }{\partial x} \right)_{ 0^+}-\left( \frac{ \partial\left[ D(x) \bar{\mathbb{P}}(x,s|0) \right] }{\partial x} \right)_{ 0^-}=-1.
\label{appen-dist-Pl-eq-9} 
\end{align}
We next insert $\bar{\mathbb{P}}(x,s|0)$ from Eq. \eqref{appen-dist-Pl-eq-81} in Eq. \eqref{appen-dist-Pl-eq-9} which results in $\mathbb{C}_7$ as
\begin{align}
\mathbb{C}_7(s) = \frac{2^{\frac{1}{2+\alpha}}}{(2+\alpha) \Gamma \left( \frac{1+\alpha}{2+\alpha}\right)} a_s^{-\frac{1}{2}}.
\label{appen-dist-Pl-eq-7}
\end{align}
Inserting this in Eq. \eqref{appen-dist-Pl-eq-81}, we obtain the exact form of $\bar{\mathbb{P}}(x,s|0)$. Since, we are interested in $\bar{\mathbb{P}}(\epsilon, s|0)$ with $\epsilon \to 0$ for last passage time [see Eq. \eqref{new-appen-last-passage-time-eq-2}], we provide below only the expression of $\bar{\mathbb{P}}(\epsilon, s|0)$:
\begin{align}
\bar{\mathbb{P}}(\epsilon, s|0) \simeq \frac{\mathcal{A}_L(\epsilon)}{s^{\frac{1}{2+\alpha}}},
\label{appen-dist-Pl-eq-8}
\end{align}
where the function $\mathcal{A}_L(\epsilon)$ is defined as
\begin{align}
\mathcal{A}_L(\epsilon) =  \frac{2^{-\frac{\alpha}{2+\alpha}}~\mathcal{D}_{\alpha}~ \Gamma \left( \frac{1}{2+\alpha} \right)}{D(\epsilon)(2+\alpha)  \Gamma \left( \frac{1+\alpha}{2+\alpha} \right)}.
\label{appen-dist-Pl-eq-9}
\end{align}
\subsection{Computation of $\bar{S}_0(s|\epsilon) $}
We now calcuate $\bar{S}_0(s|\epsilon) $ which is essential for computing $\bar{\mathcal{P}}_{ \ell } \left( p|s \right)$ in Eq. \eqref{new-appen-last-passage-time-eq-2}. To compute $\bar{S}_0(s|\epsilon) $, we proceed exactly as in Sec. \ref{appen-surv}. In order to avoid the repetition, we present only the final result here.
\begin{align}
\bar{S}_0(s|\epsilon) \simeq \frac{\mathcal{B}_L(\epsilon)}{s^{\frac{1+\alpha}{2+\alpha}}},
\label{appen-dist-Pl-eq-10}
\end{align}
where the function $\mathcal{B}_L(\epsilon)$ is defined as
\begin{align}
\mathcal{B}_L(\epsilon) = \frac{\epsilon (2+\alpha)~ \Gamma \left( \frac{1+\alpha}{2+\alpha}\right)}{\left( 4 \mathcal{D}_{\alpha} \right)^{\frac{1}{2+\alpha}}  \Gamma \left( \frac{1}{2+\alpha}\right)}.
\label{appen-dist-Pl-eq-11}
\end{align}
Plugging the forms of $\bar{\mathbb{P}}(\epsilon, s|0)$ and $\bar{S}_0(s|\epsilon)$ from Eqs. \eqref{appen-dist-Pl-eq-8} and \eqref{appen-dist-Pl-eq-10} in Eq. \eqref{new-appen-last-passage-time-eq-2} gives
\begin{align}
\bar{\mathcal{P}}_{ \ell } \left( p|s \right) \simeq \frac{\mathcal{A}_L (\epsilon)~\mathcal{B}_L (\epsilon)}{\mathcal{N}_L(\epsilon)}~\frac{1}{s^{\frac{1+\alpha}{2+\alpha}} \left(s+p \right)^{\frac{1}{2+\alpha}}}.
\label{new-appen-last-passage-time-eq-5}
\end{align}
We now have to specify the normalisation factor $\mathcal{N}_L(\epsilon)$. To evaluate this factor, we use the normalisation condition $\bar{\mathcal{P}}_{ \ell } \left( 0|s \right) = 1/s $ from which it is easy to show that $\mathcal{N}_L(\epsilon) =\mathcal{A}_L (\epsilon)~\mathcal{B}_L (\epsilon)$. This leads us to write $\bar{\mathcal{P}}_{ \ell } \left( p|s \right)$ as
\begin{align}
\bar{\mathcal{P}}_{ \ell } \left( p|s \right)= \frac{1}{s^{\frac{1+\alpha}{2+\alpha}} \left(s+p \right)^{\frac{1}{2+\alpha}}}.
\label{new-appen-last-passage-time-eq-6}
\end{align}
\greenw{Finally performing the double inverse Laplace transformation for $\alpha >-1$ gives that the distribution $\bar{\mathcal{P}}_{ \ell } \left( p|s \right)$ obeys the scaling relation} 
\begin{align}
\mathcal{P}_{ \ell } \left( t_{\ell}|t \right) = \frac{1}{t} \mathcal{G}_{\ell}^{\alpha} \left( \frac{t_{\ell}}{t}\right),
\label{new-appen-last-passage-time-eq-7}
\end{align}
with the scaling function $\mathcal{G}_{\ell}^{\alpha} \left( z \right)$ given by
\begin{align}
\mathcal{G}_{\ell}^{\alpha} \left( z \right) = \frac{z^{-\frac{1+\alpha}{2+\alpha}} \left( 1-z \right) ^{-\frac{1}{2+\alpha}}}{\Gamma \left( \frac{1+\alpha}{2+\alpha}\right)~\Gamma \left( \frac{1}{2+\alpha}\right)}.
\label{new-appen-last-passage-time-eq-9}
\end{align}